\documentclass[a4paper,article,11pt,english,amsmath,amssymb,showpacs,floatfix,notitlepage,nofootinbib,superscriptaddress]{revtex4-1}
\usepackage[dvips]{graphics,color}
\usepackage{graphicx}
\usepackage{wrapfig}
\usepackage{mathrsfs}
\usepackage{ulem}
\usepackage{bm}
\usepackage[utf8]{inputenc}
\usepackage[unicode=true,pdfusetitle,
	bookmarks=true,bookmarksnumbered=false,bookmarksopen=false,
breaklinks=false,pdfborder={0 0 1},colorlinks=false]{hyperref}
\hypersetup{colorlinks,citecolor=blue,linkcolor=blue}
\usepackage{subfigure,dcolumn}
\usepackage[T2A,T1]{fontenc}
\usepackage[russian,english]{babel}

\usepackage{listings}
\usepackage{ulem} 

\newcommand{ \be }{\begin{eqnarray}}
\newcommand{ \ee }{\end{eqnarray}}
\newcommand{ \la }{\langle}
\newcommand{ \ra }{\rangle}

\newcommand{ \mean }[1]{\la #1 \ra}

\newcommand{\La}{\langle}
\newcommand{\Ra}{\rangle}

\usepackage{color}
\definecolor{dgreen}{cmyk}{1.,0.,1.,0.2}        
\definecolor{orange}{cmyk}{0.,0.353,1.,0.}    

\newcommand{ \Num}[2]{\mathrm{N}\mean{#1}_{#2}}
\newcommand{ \Den}[2]{\mathrm{D}\mean{#1}_{#2}}

\begin{document}

\title {Collective flow and hydrodynamics in large and small systems at the LHC}

\author{Huichao Song}
\email{Huichao Song: Huichaosong@pku.edu.cn}
\affiliation{Department of Physics and State Key Laboratory of Nuclear Physics and Technology, Peking University, Beijing 100871, China}
\affiliation{Collaborative Innovation Center of Quantum Matter, Beijing 100871, China}
\affiliation{Center for High Energy Physics, Peking University, Beijing 100871, China}

`
\author{You Zhou}
\email{You Zhou: you.zhou@cern.ch}
\affiliation{Niels Bohr Institute, University of Copenhagen, Blegdamsvej 17, 2100 Copenhagen, Denmark}

\author{Katar\'{i}na~Gajdo\v{s}ov\'{a}}
\affiliation{Niels Bohr Institute, University of Copenhagen, Blegdamsvej 17, 2100 Copenhagen, Denmark}

\begin{abstract}
\vspace{0.5 cm}
In this article, we briefly review the recent progress on collective flow and hydrodynamics in large and small systems at the Large Hadron Collider (LHC), which includes the following topics: extracting the QGP viscosity from the flow data, initial state fluctuations and final state correlations at 2.76 A TeV Pb--Pb collisions, correlations and collective flow in high energy p--Pb and p--p collisions.
\end{abstract}


\maketitle

\section{Introduction}

At extremely high temperatures and densities, the strong-interaction matter can experience a phase transition
and form a hot and thermalized medium called the quark-gluon plasma (QGP), where quarks and gluons are no longer confined, but propagate over larger distances than the typical size of a hadron~\cite{Lee:1974ma,Collins:1974ky}. Around a few microseconds after the Big Bang, the QGP once filled in the whole early universe. With the expansion and cooling down of the universe, the primordial QGP went through a phase transition and formed hadrons, including protons and neurons, the basic building blocks of our current visible word. The QGP can also be created at the Relativistic Heavy-Ion Collider (RHIC) and the Large Hadron Collider (LHC), where the ulta-relativistic collisions of heavy ions allow us to achieve the needed extreme conditions for the QCD phase transitions and for the formation of the QGP~\cite{Lee:1974ma,Collins:1974ky,Baumgardt:1975qv}.

Since the running of RHIC in 2000, strong evidences were gradually accumulated for the creation of the QGP in the high energy nucleus-nucleus collisions~\cite{Arsene:2004fa,Back:2004je,Adams:2005dq,Adcox:2004mh,Arsene:2004fa,Gyulassy:2004vg,Muller:2006ee}. The observation of strong collective flow and the successful descriptions from hydrodynamics reveal that the QGP is a strongly-coupled system and behaves like an almost perfect liquid~\cite{Gyulassy:2004vg,Muller:2006ee,Kolb:2003dz}. It was also realized that, since the nucleons inside the colliding nuclei constantly change their positions, the created QGP fireballs fluctuate event-by-event~\cite{Alver:2006wh,Miller:2003kd,Alver:2008zza}. The collective expansion of the hot systems transforms the initial spacial inhomogeneities and deformation into anisotropic momentum distributions of final produced particles~\cite{Ollitrault:1992bk,Voloshin:1994mz}, which can be quantitatively evaluated by various flow observables~\cite{Voloshin:2008dg,Snellings:2011sz,Heinz:2013th,Gale:2013da,Song:2013gia,Luzum:2013yya,Jia:2014jca}.
For example, the elliptic flow  $v_2$ is associated with the elliptic deformation of the initial fireball,  the triangular flow
$v_3$  is mainly controlled by the event-by-event fluctuations of the systems and the quadrangular flow $v_4$ is driven by both initial spacial deformations and inhomogeneities of the created fireball, etc~\cite{Alver:2010dn,ALICE:2011ab,Gardim:2011xv,ATLAS:2012at}. Besides these individual flow harmonics, other flow observables, such as such as $v_n$ in ultra-central collisions~\cite{Luzum:2012wu,CMS:2013bza}
, the distributions of event-by-event flow harmonics~\cite{Aad:2013xma,Gale:2012rq}, the event-plane correlations~\cite{Aad:2014fla,Qiu:2012uy}, and the correlations between different flow harmonics~\cite{Aad:2015lwa,ALICE:2016kpq,Giacalone:2016afq,Zhu:2016puf,Qian:2016pau}, the de-correlation of the flow vector~\cite{Heinz:2013bua,Gardim:2012im,Khachatryan:2015oea}, etc., have also been intensively measured and studied in the high energy Pb--Pb collisions at the LHC. Together with the sophisticated event-by-event simulations from hydrodynamics and hybrid models, these different flow observables provide important information on the properties of the QGP fireball and help to constrain the the initial conditions of the colliding systems~\cite{Voloshin:2008dg,Snellings:2011sz,Heinz:2013th,Luzum:2013yya,Gale:2013da,Jia:2014jca,
Song:2013gia}.

The measurements of the azimuthal correlations in small systems, e.g. in p--Pb and p--p collisions at the LHC, were originally aimed to provide the reference data for the high-energy nucleus-nucleus collisions. However, lots of unexpected phenomena were discovered in experiments, which indicates the development of collective flow in the small systems. As the collision energy increased to the LHC regime, the multiplicities in "ultra-central" p--Pb and p--p collisions is comparable to the ones in peripheral Pb--Pb collisions, where that final state interactions become possibly sufficient to develop the collective expansion.  A comparison of the
two particle correlations in high multiplicity p--Pb collisions at $\sqrt{s_{\rm NN}}=$ 5.02 TeV and in peripheral Pb--Pb collisions at $\sqrt{s_{\rm NN}}=$ 2.76 TeV show a surprisingly similar correlation structures for these events with similar multiplicity cuts ~\cite{CMS:2012qk,Abelev:2012ola,Aad:2013fja,Khachatryan:2015waa}. Besides, a changing sign of the 4-particle cumulants~\cite{Aad:2013fja,Abelev:2014mda,Khachatryan:2015waa} and a $v_{2}$ mass mass-ordering feature of identified hadrons~\cite{ABELEV:2013wsa,Khachatryan:2014jra} and other flow-like signals have also been observed in the high multiplicity p-Pb collisions.  The related hydrodynamic simulations have successfully reproduced many of these experimental data, which strongly support the observation of collective flow in high-multiplicity p--Pb collisions~\cite{Bozek:2011if,Bozek:2012gr,Bozek:2011if,Bozek:2013ska,Bzdak:2013zma,Qin:2013bha,Werner:2013ipa,Schenke:2014zha}. For p--p collisions at $\sqrt{s_{\rm NN}}=$ 7 TeV and 13 TeV,  similar results, but with smaller magnitudes, have been obtained for many of these flow-like observables~\cite{Khachatryan:2010gv,Li:2012hc,Aad:2015gqa,Khachatryan:2015lva,Khachatryan:2016txc,Dusling:2015gta}. Although these measurements may associated with the collective expansion in the small p--p systems, more detailed investigations are still needed to further understand of the physics behind.

In this paper, we will review the recent progress on collective flow and hydrodynamics in large and small systems at the LHC. In Sec.~II and Sec.~III, we will introduce hydrodynamics, hybrid models and flow measurements. In Sec.~IV, we will review recent progress on extracting the QGP viscosity from the flow data at the LHC. In Sec.~V, we will focuses on initial state fluctuations and final state correlates in Pb--Pb collisions at 2.76 A TeV. In Sec.~VI, we will review the correlations and collective flow in small systems.  Sec.~VII will briefly summarize and conclude this paper.

\section{Hydrodynamics and hybrid model}

\subsection{Viscous hydrodynamics}

\quad Viscous hydrodynamics is a widely used tool to describe the expansion of the QGP fireball and to study the soft hadron data for the heavy ion collisions at RHIC and the LHC~\cite{Teaney:2009qa,Romatschke:2009im,Huovinen:2013wma,Heinz:2013th,Gale:2013da,Song:2013gia,Romatschke:2007mq,Luzum:2008cw,
Song:2007fn,Song:2007ux,Song:2009gc,Dusling:2007gi,Molnar:2008xj,Bozek:2009dw,Chaudhuri:2009hj,
Schenke:2010rr}. It solves the transport equations of energy momentum tensor and net charge current, which are written as
\begin{subequations}
\begin{eqnarray}
 && \partial_\mu T^{\mu \nu}(x)=0, \\
 && \partial_\mu N^{\mu} (x)=0\, .   
\end{eqnarray}
\end{subequations}
If the systems are very close to local equilibrium, the energy momentum tensor and the net baryon charge current can be decomposed as: $T^{\mu \nu}=(e+p) u^{\mu}u^{\nu}-p g^{\mu\nu}$ and $N^{\mu}=nu^{\mu}$. Therefore, the fourteen variables in $T^{\mu\nu}$ and $N^{\mu}$  are reduced to six independent unknowns: the energy density $e$, pressure $p$ and net baryon density $n$, and 3 independent components in the four velocity $u^\mu$. The relativistic hydrodynamics are then simplified as ideal hydrodynamics. With an additional input, the equation of state (EoS) $p=p(n,e)$, and the chosen initial and final conditions, the ideal hydrodynamic equations can be numerically solved to simulate the evolution of the bulk matter for the relativistic heavy ion collisions~\cite{Kolb:2003dz}.

For a near equilibrium system, one need to implement the relativistic viscous hydrodynamics (or the so-called relativistic dissipative fluid dynamics). In the Landau frame, $T^{\mu\nu}$ and $N^{\mu}$ are expressed as: $T^{\mu \nu}=(e+p+\Pi) u^{\mu}u^{\nu}-(p +\Pi)g^{\mu\nu}+\pi^{\mu \nu}$, $N^{\mu}=nu^{\mu}-\frac{n}{e+p}q^{\mu} $.  Here, $\pi^{\mu \nu}$ is the shear stress tensor, $\Pi$ is the bulk pressure and $q^{\mu}$ is the heat flow. From the 2nd law of thermal dynamics or from the kinetic theory, one could obtain the viscous equations of $\pi^{\mu \nu}$, $\Pi$ and $q^\mu$, which are expressed as~\cite{Israel:1976tn,Muronga:2004sf}:
\begin{subequations}
\begin{eqnarray}
&&\Delta ^{\mu \alpha} \Delta ^{\nu \beta}\dot{ \pi}_{\alpha \beta}
=-\frac{1}{\tau_{\pi}}\bigg[\pi^{\mu\nu}-2\eta \nabla ^{\La \mu}u
^{\nu\Ra}-l_{\pi q} \nabla^{\La\mu} q^{\nu\Ra}
  + \pi_{\mu\nu} \eta T
\partial_{\alpha} \big( \frac{\tau_\pi u^{\alpha}}{2 \eta T} \big)
\bigg], \ \ \ \\
&&\dot{\Pi}=-\frac{1}{\tau_{\Pi}}\bigg[\Pi+\zeta \theta -l_{\Pi q}
\nabla_{\mu} q^{\mu}+\Pi \zeta T
\partial_{\mu} \big( \frac{\tau_\Pi u^{\mu}}{2\zeta T} \big) \bigg],
\quad  \\
&&\Delta^{\mu}_{\nu}\dot{q}^{\nu}=-\frac{1}{\tau_{q}}\bigg[q_{\mu}+\lambda
\frac{n T^2}{e+p}\nabla ^{\mu}\frac{\nu}{T} + l_{q \pi}\nabla_\nu
\pi^{\mu\nu}
 + l_{q \Pi}\nabla^\mu \Pi - \lambda T^2 q^{\mu}
\partial_{\mu}\big( \frac{\tau_q u^{\mu}}{2\lambda T^2}
\big)\bigg],
\label{dissi-trans-b}
\end{eqnarray}
\end{subequations}
where $\Delta^{\mu\nu} =g^{\mu\nu}{-}u^\mu u^\nu$,
$\nabla ^{\La \mu}u ^{\nu\Ra}=\frac{1}{2}(\nabla^\mu u^\nu+\nabla^\nu
u^\mu)-\frac{1}{3}\Delta^{\mu\nu}\partial_\alpha u^\alpha$ and
$\theta=\partial \cdot u$. $\eta$ is the shear viscosity, $\zeta$ is the bulk viscosity, $\lambda$ is the heat conductivity,
and $\tau_{\pi}$, $\tau_{\Pi}$ and  $\tau_{q}$ are the corresponding relaxation times.

The above Israel-Stewart formalism can also be obtained from the kinetic
theory~\cite{Baier:2006um,Baier:2007ix,Betz:2008me,Denicol:2012cn,Denicol:2012es} or from
the conformal symmetry constraints~\cite{Baier:2007ix}~\footnote{The traditional 2nd order viscous hydrodynamics works for a near equilibrium system with isotropic momentum distributions. It can not apply to an anisotropic system at very early time~\cite{Martinez:2010sc,Florkowski:2010cf,Jeon:2016uym} or a correlated fluctuating system near the QCD critical point~\cite{Stephanov:2008qz,Stephanov:2011pb,Jiang:2015hri,Jiang:2015cnt} where the traditional expansion of the microscopic distribution function fails.  For the recent development on anisotropic hydrodynamics or chiral hydrodynamics, please refer to~\cite{Martinez:2010sc,Florkowski:2010cf,Jeon:2016uym,Martinez:2012tu,Florkowski:2013lza,Ryblewski:2012rr,
Bazow:2013ifa,Bazow:2015cha,Strickland:2016ezq} and~\cite{Paech:2003fe,Nahrgang:2011mg,Nahrgang:2011mv,Herold:2013bi,Herold:2016uvv}.}.
These different derivations give different higher order terms for the 2nd order viscous equations. In general, the contributions of the higher order terms are pretty small or even negligible for a hydrodynamic evolution with small shear and bulk viscosity, which will will not significantly influences the final flow observables~\footnote{Note that, to obtain a good agreement with the microscopic kinetic theory, a proper resummation of the irreducible moments is essential
for the computation of the transport coefficients, especially for a fluid-dynamics with heat flow included. Please refer to~\cite{Denicol:2012vq} for details. }.

\vspace{0.2cm}
\underline{\emph{The equations of state (EoS)}}:
\vspace{0.10cm}

Besides these hydrodynamic equations, one needs to input an EoS to close the system for the numerical simulations or analytical solutions. Currently, many groups use a state-of-the-Art EoS, called s95p-PCE,  which combines a parameterized/tablated lattice EoS for the baryon free QGP phase with a hadronic EoS with effective chemical potentials for
the partially chemical equilibrium hadronic phase~\cite{Huovinen:2009yb,Shen:2010uy}.
Ref.~\cite{Huovinen:2009yb} also compared the hydrodynamic calculations using various equations of state constructed with different speed of sound, which found that the spectra and elliptic flow are only slightly influenced by the inputting EoS. The main uncertainties of the hydrodynamic calculations come from the initial conditions, which will be introduced and discussed below.

\vspace{0.2cm}
\underline{\emph{Initial conditions}}:
\vspace{0.10cm}


The initial condition is a necessary input for the hydrodynamic simulations. As an open issue related to the creation and thermalization of the QGP, it brings some uncertainties, more or less, for many flow observables in the hydrodynamic calculations. There are many types of initial condition models developed by different groups.  The traditional {\tt Glauber model} assumes zero transverse velocity at the starting time and constructs the initial entropy/energy density profiles from a mixture of the wounded nucleon and binary collision densities~\cite{Kolb:2000sd}. The {\tt KLN} model treat the degrees of freedom of the initial systems as gluons and calculate the initial density profiles from the $k_T$ factorization formula~\cite{Kharzeev:2000ph}. In the later developed Monte-Carlo versions, called ({\tt MC-Glauber} and {\tt MC-KLN})~\cite{Miller:2007ri,Drescher:2006ca,Hirano:2009ah}, the event-by-event fluctuations are built through the positions fluctuations of individual nucleons inside each colliding nuclei. For the {\tt AMPT} initial conditions, the initial profiles are constructed from the energy and momentum decompositions of individual partons, which fluctuate in both momentum and position space~\cite{Bhalerao:2015iya,Pang:2012he,Xu:2016hmp}. With an additional Gaussian smearing factor, the fluctuation scales related to the energy decompositions become changeable, which helps to balance the initial eccentricities at different order. As a successful initial condition model, {\tt IP-Glasma}~\cite{Schenke:2012fw} includes both the nucleon position fluctuations and the color charge fluctuations. It uses the IP-Sat model to generate the wave-functions of high energy nuclei/nucleon and then implements a classical Yang-Mills dynamics to simulate the pre-equilibrium evolution of the early glasma stage. Another successful initial condition model in {\tt EKRT}~\cite{Paatelainen:2013eea,Niemi:2015qia} combines the PQCD minijet production with the gluon saturation and generates the energy density profiles after a pre-equilibrium Bjorken free streaming. The recently developed {\tt T\raisebox{-.5ex}{R}ENTo} model~\cite{Moreland:2014oya} is a parametric initial condition model based on the eikonal entropy deposition via a reduced thickness function. With an introduced entropy deposition parameter, {\tt{T\raisebox{-.5ex}{R}ENTo}} model could reproduce the initial eccentricities of various initial condition models that belong to different classes, such as {\tt MC-KLN}, {\tt MC-Glauber}, {\tt EKRT}, {\tt IP-Glasma} and etc..

Many initial condition models neglect the initial flow from the pre-equilibrium stage. Recently, the effects of pre-equilibrium evolution have been estimated in Ref~\cite{Liu:2015nwa} through evolving the free streaming particles from MC-Glauber and MC-KLN models, which demonstrated that such pre-equilibrium dynamics significantly increases the initial flow and reduces the initial spacial eccentricities. More sophisticated investigations on pre-equilibrium dynamics can be, in principle, carried on within the framework of dynamical models like {\tt EPOS}~\cite{Werner:2010ny}, {\tt AMPT}~\cite{Bhalerao:2015iya,Pang:2012he,Xu:2016hmp},  {\tt EKRT}~\cite{Paatelainen:2013eea,Niemi:2015qia}, {\tt IP-Glasma}~\cite{Schenke:2012fw}, {\tt URQMD}~\cite{Petersen:2009vx,Petersen:1900zz} and etc.. After matching the energy-momentum tensor at a switching point, one could principally obtained 3+1-d fluctuating profiles of initial energy density and initial flow for the succeeding hydrodynamic simulations.  However, many past studies focus on the initial state fluctuations on the transverse plane, which neglect the fluctuation patterns along the longitudinal direction. The  AMPT + ideal hydrodynamic simulations~\cite{Pang:2012he} demonstrate that evolving early hot spots in the longitudinal directions could dissipate part of the transverse energy, which leads to a suppression of the final flow anisotropy. Recently, the IP-Glasma model has been extended to three dimension with the explicit small x evolutions of the gluon distributions~\cite{Schenke:2016ksl}. Although the related energy momentum tensors can be in principle used in the succeeding hydrodynamic simulations, additional works are still required to further extend the distributions to the large rapidity regime with the consideration of large x effects.

\vspace{0.2cm}
\underline{\emph{Freeze-out / decoupling }}:
\vspace{0.10cm}

Pure hydrodynamic simulations assume free-streaming hadrons directly emit from a decoupling surface defined by a constant temperature or energy density or other kinetic variables\cite{Kolb:2003dz,Teaney:2009qa}.  The momentum distributions of various emitted thermal hadrons can be calculated  with the Cooper-Frye formula~\cite{Cooper-Frye} using the freeze-out information on the freeze-out surface (For the details of the Cooper-Frye formula, please refer to~\cite{Cooper-Frye,Kolb:2003dz} as well as the following Section II. B for details).  With the corresponding decay channels, the unstable hadron resonances delay into stable ones with some momentum distributions that can be further analyzed to compare with the experimental data.  In the constant temperature decoupling scenario, the decoupling temperature $T_{dec}$
strongly depends on the EoS and other hydrodynamic inputs. For s95p-PCE, $T_{dec}$ is generally set to 100-120 MeV in order to fit the mean $p_T$ of various hadrons with a sufficient build up of the radial flow~\cite{Kolb:2003dz,Shen:2010uy}.

\subsection{Hybrid models}
\quad A hybrid model matches the hydrodynamic description of the QGP fluid to a hadron cascade simulation for the evolution of the hadron resonance gas at a switching temperature near $T_c$. The early ideal hydrodynamics + hadron
cascade hybrid model simulations have showed the hadronic matter is highly viscous, which largely suppress the elliptic flow when compared with the pure hydrodynamic calculations with a partially chemical equilibrium EoS~\cite{Hirano:2005wx}. Motivated by this, different groups have extended 2+1-d or 3+1-d viscous hydrodynamics with a hadronic afterburner~\cite{Song:2010aq,Ryu:2012at,Karpenko:2013ama}. Such hybrid models give a more realistic description for the hadronic evolution of the QGP fireball, which also naturally imprint the off-equilibrium chemical and thermal freeze-out procedures of various hadron species.

The key component of a hybrid model is the particle event generator that convert the hydrodynamic
output on the switching hyper surface into various hadrons for the succeeding hadron cascade simulations.
More specifically, such Monte Carlo event generator produces particles
with specific momentum and position information according to the differential Cooper-Frye
formula~\cite{Song:2010aq}:
\begin{eqnarray}
  E\frac{d^3N_i}{d^3p}(x) &=& \frac{g_i}{(2\pi)^3}
  p\cdot d^3\sigma(x)\, f_i(x,p)
\end{eqnarray}
Where $f_i(x,p)$ is the distribution function of hadron species $i$,
$g_i$ is the corresponding degeneracy factor and
$d^3\sigma_\mu(x)$ is a surface element on the hyper-surface
$\Sigma$, e.g., defined by a constant switching temperature $T_{sw}$.
Generally, the switching temperature $T_{sw}$ is set to around 160 MeV, 
which is
close to the phase transition temperature of the QCD matter at zero
chemical chemical potential~\cite{Ding:2015ona}.  For a viscous QGP fluid,
the distribution function $f(x,p)$ include an ideal part and an off-equilibrium part
$f=f_0+\delta f$, where $\delta f$ generally takes the form: 
$\delta f=f_0 \bigl(1{\mp}f_0\bigr)\frac{p^\mu p^\nu \pi_{\mu\nu}}{2T^2\left(e{+}p\right)}$~\cite{
Romatschke:2007mq,Luzum:2008cw,Song:2007fn,Song:2007ux,Song:2009gc,Dusling:2007gi}
~\footnote{The full off-equilibrium distribution includes the contributions from shear stress tensor, bulk pressure and heat flow: $\delta f=\delta f_{shear}+ \delta f_{bulk}+ \delta f_{heat}.$  For the bulk viscous correction, there
are different proposed forms of $\delta f_{bulk}$~\cite{Dusling:2011fd,Noronha-Hostler:2013gga}, which brings certain amount of uncertainties for some related flow observables.  Considering this complicity as well as the negligible heat conductivity, one generally takes this simple form of  $\delta f$  with only shear viscous correction  for the viscous hydrodynamics and hybrid model calculations at top RHIC and the LHC energies.}.

After converting the fluid into many individual hadrons of various species, the hybrid model implement a hadron cascade model to simulate the
microscopic evolution of the hadron resonance gas. The hadron cascade model, for example,
Ultra-relativistic Quantum Molecular Dynamics (UrQMD)~\cite{Bass:1998ca,Bleicher:1999xi} solves a large set of
Boltzmann equations for various hadron species:
\begin{eqnarray}
 \frac{d f_i(x,p)}{dt}= C_i(x,p)
\end{eqnarray}
where $f_i(x,p)$ is the distribution function and $C_i(x,p)$ is the collision terms for hadron
species i. With such equations, the hadron cascade model propagate various hadrons with
classical trajectories, together with the elastic scatterings, inelastic scatterings and resonance decays.
After all the collisions and decays cease, the system stops evolution and outputs the information
of produced hadron which can be further analyzed to compared with the experimental data~\cite{Bass:1998ca,Bleicher:1999xi}.

Compared with hydrodynamic calculations, the hybrid model improves the description of the
hadronic evolutions and the decoupling procedure, which leads to a nice descriptions of the flow harmonics of identified hadrons, especially for the mass-splitting between pions and protons~\cite{Song:2013qma,Heinz:2011kt}. Meanwhile, the
imprinted baryon-antibaryon annihilations in the hadronic cascade sector also largely reduce the production of
proton and antiproton, which helps to achieve a nice fit of particle yields of various
hadron species~\cite{Song:2013qma,Song:2012ua}.

\vspace{0.2cm}
\underline{\emph{2+1-d vs 3+1-d model}}:
\vspace{0.10cm}

For hydrodynamics or hybrid models, the 2+1-d simulations with a longitudinal boost invariance are  more
computational efficient than the full 3+1-d simulations. Before 2010, many developed viscous hydrodynamic codes  are (2+1)-dimensional using the Bjorken approximation~\cite{Romatschke:2007mq,Luzum:2008cw,Song:2007fn,Song:2007ux,Song:2009gc,Dusling:2007gi,
Molnar:2008xj,Bozek:2009dw,Chaudhuri:2009hj}. The published VISHNU code is also basically a (2+1)-d hybrid code since it implements the (2+1)-d viscous hydrodynamic simulations for the evolution of the QGP phase. Although the succeeding UrQMD afterburner are (3+1)-dimensional, the longitudinal boost invariance are still approximately conserved at mid-rapidity after the hadronic evolution~\cite{Song:2010aq}.  Recently, several groups~\cite{Schenke:2010rr,Bozek:2011ua,Vredevoogd:2012ui,Nonaka:2013uaa,DelZanna:2013eua,Karpenko:2013wva} further developed the full (3+1)-d viscous hydrodynamics or hybrid models without a longitudinal boost invariance. Such full (3+1)-d simulations
could provide full space-time evolution profiles for the EM and hard probes. They can also be widely used to investigate the longitudinal fluctuations, to study the physics for asymmetric collision systems, such as p+Pb, d+Au and Cu+Au, etc, and to provide more realistic calculations / predictions for the heavy ion collisions at lower collision energies.

\subsection{Event-by-event simulations}

\quad As introduced in Sec.II A, the initial profiles of the created QGP fireball fluctuate event-by-event, which leads to
the final state correlations and collective flow for the nucleus-nucleus collisions at RHIC and the LHC~\cite{Alver:2006wh,Miller:2003kd,Alver:2008zza}. For computational efficiency, the early hydrodynamics or hybrid model simulations input smooth initial profiles obtained through averaging a large number of events generated from some specific fluctuating initial conditions and then implement the so-called \texttt{single-shot simulations}. An alternative approach is the \texttt{event-by-event simulations}, which simultaneously run a large number of  simulations with the input of individually fluctuating initial profiles. Past research has showed, due to the the approximate linear hydrodynamic response $v_2 \propto \varepsilon_2$ and $v_3 \propto \varepsilon_3$,  the elliptic and triangular flow can be nicely described by the single-shot hydrodynamic simulations with properly chosen initial conditions and well tuned parameter sets.
However, such the single shot simulation fails to describe other higher order flow harmonics due to the mode coupling effects. Furthermore, some flow observables, such as event-by-event flow harmonics~\cite{Aad:2013xma,Gale:2012rq}, the event-plane correlations~\cite{Aad:2014fla,Qiu:2012uy}, and the correlations between different flow harmonics~\cite{Aad:2015lwa,Qian:2016pau,ALICE:2016kpq,Zhu:2016puf,Giacalone:2016afq}, etc., can not be directly calculated by the single-shot hydrodynamics or hybrid models, which requires to implement the event-by-event simulations (please also refer to Sec. V for details).

Since 2010, many groups have developed the event-by-event hydrodynamics / hybrid models to study the initial fluctuations, hydrodynamic response and the corresponding final state correlations~\cite{Schenke:2012fw,Gale:2012rq,Pang:2012he,Petersen:2010cw,Qin:2010pf,Holopainen:2010gz,Qiu:2011iv,Qiu:2012uy,
Shen:2014vra,Paatelainen:2013eea,Niemi:2015qia,Bhalerao:2015iya,Xu:2016hmp}.
In general, such event-by-event simulations is computational expansive. For example, the iEBE-VISHNU simulations for the
correlations between flow harmonics have used 30000 CPU hours in Tianhe-1A  National Supercomputing Center in Tianjin China.
Recently, the OSU-Kent group has developed the massively parallel simulations for 3+1-d viscous hydrodynamics on graphics processing units with CUDA and demonstrated that such GPU simulations are approximately two orders of magnitude faster than the corresponding simulations from CPU~\cite{Bazow:2016yra}. With the development of computer science and the reduced cost of GPU, the GPU-based simulations will become a popular trend for the massive hydrodynamic calculations in the future.

\section{Flow method}
\label{sec:method}
The anisotropic flow evaluates the anisotropy in particle momentum distributions correlated with the
flow symmetry plane $\Psi_{n}$~\cite{Ollitrault:1992bk}.  The various characteristic  patterns of the anisotropic flow can be obtained from a Fourier expansion of the event averaged azimuthal particle distribution~\cite{Voloshin:1994mz}:
\begin{equation}
\frac{{\rm d} N}{{\rm d} \varphi} \propto 1+ 2 \sum_{n=1}^{\infty} v_{n} \, e^{in(\varphi-\Psi_{n})}
\end{equation}
where $v_{n} = \langle cos \, n(\varphi - \Psi_n) \rangle$ is anisotropic flow and $\Psi_{n}$ is the corresponding flow symmetry plane.

Since the flow symmetry plane is not a direct observable, the anisotropic flow $v_{n}$ can not
be measured directly. A popular approach is the event-plane method~\cite{Poskanzer:1998yz}, which has been widely used to calculate the azimuthal correlation of emitted particles with respect to the event-plane. However, it was found that the results from event-plane method strongly depends on the resolution of the event-plane, which introduces an uncontrolled bias in the measurement~\cite{Luzum:2012da}.
As an alternative approach, the multi-particle azimuthal correlations method~\cite{Bilandzic:2010jr,Bilandzic:2013kga} has been developed and improved in the past ten years, which allows an unambiguous measurement of the underlying anisotropic flow and eliminates the detector bias.

\vspace{0.2cm}
\underline{\emph{2- and multi-particle correlations}}
\vspace{0.10cm}

Azimuthal correlations of 2 or multi-particles are calculated in two steps~\cite{Bilandzic:2010jr,Bilandzic:2013kga}. First, one obtains an average over all particles in a single-event, and then calculate an average over all events.  The single-event 2-particle correlation is defined as:
\begin{eqnarray}
\langle cos \,n(\varphi_1 - \varphi_2) \rangle = \langle e^{in(\varphi_1 - \varphi_2)} \rangle \qquad \qquad \qquad \qquad \qquad
\label{eq:2pc}
\end{eqnarray}
Here, $\langle ... \rangle$ denotes an average over all particles in a single-event.
An average of the 2-particle correlation over all events is generally denoted by $\langle \langle  ... \rangle \rangle= \langle \langle e^{in(\varphi_1 - \varphi_2)} \rangle \rangle$. Such correlations can serve as an estimate of the flow harmonics $v_n$ without the knowledge of the symmetry plane, which can also be demonstrated as:
\begin{eqnarray}
\qquad \langle \langle e^{in(\varphi_1 - \varphi_2)} \rangle \rangle  &=& \langle \langle e^{in(\varphi_1 - \Psi_n - \varphi_2 + \Psi_n)} \rangle \rangle   \nonumber \\
&=&\langle \langle e^{in(\varphi_1 - \Psi_n)} \rangle \langle e^{in(\varphi_2 - \Psi_n)}  \rangle + \delta_n \rangle \approx  \langle v_n^2 \rangle + \delta_n, \quad
\label{eq:22-pc}
\end{eqnarray}
where $\delta_n$ is called non-flow. It is a term related to the statistical fluctuations, which implies that $\langle AB \rangle \neq \langle A \rangle \langle B \rangle$, or originated from the 2-particle correlations that is not associated with the collective expansion~\cite{Snellings:2011sz}.

The formulas above can be extended to a generic notation for the average single-event k-particle correlators with mixed harmonics:
\begin{eqnarray}
\langle \cos(n_1\varphi_1\! + \!n_2\varphi_2\!+\!\cdots\!+\!n_i\varphi_i) \rangle  \,(n_1\geq n_2 \geq \cdots \geq n_i)
\qquad \qquad
\label{eq:mm-pc}
\end{eqnarray}
Here, the azimuthal angle $\varphi_i$ belongs to the reconstructed particle $i$. The self-correlations should be removed completely and only genuine multi-particle correlations left. For simplicity, we also denote this $k$-particle correlators as $\langle k \rangle_{n_{1}, n_{2}, ..., n_{k}}$ in the following context. As the case for the 2-particle correlator,
the subsequent average over all events can obtained in a similar way described in Eqs.~(\ref{eq:22-pc}). For details, please refer to~\cite{Bhalerao:2011yg}.

Calculations for the single event averaged multi-particle correlators require a large amount of the computational resources, which significantly increases for higher order correlations. A successful way to calculate these correlators in a single loop over particles in one event can be achieved by the Q-vectors, which will be introduced in the following text.

\vspace{0.2cm}
\underline{\emph{Q-cumulant method}}
\vspace{0.10cm}

In the Q-Cumulant method~\cite{Bilandzic:2010jr}, the single-event averaged correlations are calculated in terms of a $Q_{n}$-vector, which is defined as:
\begin{equation}
Q_{n} \equiv \sum_{i=1}^M  e^{in\phi_i}\,, \qquad \qquad \qquad \qquad
\label{a:Qvector}
\end{equation}
where $M$ is the number of particles in a specific event, and $\phi_i$ is the azimuthal angle of the $i$-th particle.
For azimuthal correlations involving only single harmonic, the single-event average 2-, and 4-particle azimuthal correlations can be calculated as:
\begin{eqnarray}
&&\langle 2 \rangle_{n,-n}  = \frac{|Q_{n}|^{2} - M} {M(M-1)} \label{Eq:2pc}\\
&&\langle 4  \rangle_{n,n,-n,-n} = \frac{ \, |Q_{n}|^{4} + |Q_{2n}|^{2} - 2 \cdot {\rm{Re}} \left( Q_{2n}Q_{n}^{*}Q_{n}^{*} \right)- 2 [ 2(M-2) \cdot |Q_{n}|^{2} - M(M-3) ] \, } {[ M(M-1)(M-2)(M-3) ]} \nonumber.
\end{eqnarray}
After averaging the correlators over whole event sample, one obtains the 2- and 4-particle cumulants:
\begin{eqnarray}
c_{n}\{2\} = \langle \langle 2  \rangle \rangle_{n,-n}, \qquad
c_{n}\{4\} = \langle \langle 4  \rangle \rangle_{n,n,-n,-n} - 2 \, \langle \langle 2  \rangle \rangle_{n,-n}^{2}. \qquad \qquad
\label{Eq:c46}
\end{eqnarray}
Eventually, the 2- and 4-particle and reference (integrated) flow harmonics can be calculated as:
\begin{eqnarray}
v_n\{2\} &= \sqrt{c_n\{2\}}, \qquad   v_n\{4\} &= \sqrt[4]{-c_n\{4\}}. \qquad \qquad \qquad \qquad \qquad \qquad
\end{eqnarray}

The differential flow harmonics for identified or all charged hadrons can be obtained from a single-event correlators averaged over only these particles of interest within an event. For the limitation of space, we will not further outline the lengthy formula, but refer to~\cite{Bilandzic:2010jr} for details.

As pointed out above, the non-flow effects, originated from resonance decays, jets and etc., could strongly influence the calculated flow harmonics, especially for the ones obtained from 2-particle correlations. In order to largely suppress the non-flow contribution, a successful method of applying an $|\Delta \eta|$ gap to 2-particle correlations has been developed. In this method, an analyzed event is divided into 2 sub-events with certain $|\Delta \eta|$ separation. After obtained the Q-vectors for each sub-event separately, the single-event average 2-particle correlation with $|\Delta \eta|$ gap can be calculated as:
\begin{equation}
\langle 2 \rangle_{n,-n}^{|\Delta \eta|}  = \frac{Q_{n}^{A}\cdot Q_{n}^{B*}} {M^{A}\cdot M_{B}}, \qquad \qquad \qquad \qquad
\end{equation}
where $A$ and $B$ denote two different sub-events. The corresponding final flow harmonics are usually denoted as: $v_n\{2, |\Delta \eta| > X\}$, which can be obtained in the same way as the above reference flow without $|\Delta \eta|$ gap.

\vspace{0.2cm}
\underline{\emph{Generic framework}}
\vspace{0.10cm}

In 2013, a generic framework was developed~\cite{Bilandzic:2013kga}, which enables exact and efficient evaluation of all multi-particle azimuthal correlations. This framework can be used along with a correction framework for systematic biases in anisotropic flow analyses due to the Non-Uniform Acceptance (NUA) and Non-Uniform Efficiency (NUE) effects.
For an event with multiplicity $M$, it was proposed to construct two sets for azimuthal angles of the particles $\{\varphi_1,\varphi_2,\ldots,\varphi_M \}$ and for the the corresponding weights $\{w_1,w_2,\ldots,w_M \}$.
With these two sets, one can calculate the weighted $Q_{n}$-vectors in each event, which is defined as:
\begin{equation}
Q_{n,p} \equiv \sum_{i=1}^{M}w_i^p\,e^{in\varphi_i} \,. \qquad \qquad \qquad \qquad
\label{eq:Qvector}
\end{equation}
where $w_i$ is the weight and p is the power of the weight. Correspondingly, the $i$-particle correlator is defined as:
\begin{eqnarray}
&&\Num{m}{n_1,n_2,\ldots,n_m}\equiv
\displaystyle\sum_{\begin{subarray}{c}i_1,i_2,\ldots,i_m=1\\i_1\neq i_2\neq \ldots\neq i_m\end{subarray}}^{M}\!\!\!\!\!w_{i_1}w_{i_2}\cdots w_{i_m}\,e^{i(n_1\varphi_{i_1}+n_2\varphi_{i_2}+\cdots+n_m\varphi_{i_m})}
\label{eq:num}
\end{eqnarray}
Here, the $i$-particle correlator is denoted as $\Num{m}{n_1,n_2,\ldots,n_m}$ for convenience. One could also introduce a shortcut $\Den{m}{n_1,n_2,\ldots,n_m}= \Num{m}{0,0,\ldots,0}$ and then calculate the single-event average of multi-particle azimuthal correlations via:
\begin{equation}
\langle m \rangle_{n_1,n_2,\ldots,n_m} = \frac{\Num{m}{n_1,n_2,\ldots,n_m}}{\Den{m}{n_1,n_2,\ldots,n_m}}.
\qquad \qquad \qquad \qquad \qquad
\label{eq:MPCGF}
\end{equation}

Based on this generic framework, one could explicitly outline the results for the 2- and 4-particle correlators, which can be analytically expressed in terms of the $Q_{n,p}$-vectors defined in the above context.  The single-even average 2- and 4-particle correlations could be then calculated as:  
\begin{eqnarray}
&&\langle 2 \rangle_{n_{1}, n_{2}} = \frac{\Num{2}{n_1,n_2}}{\Den{2}{n_1,n_2}}, \qquad
\langle 4 \rangle_{n_{1}, n_{2}, n_{3}, n_{4}} = \frac{\Num{4}{n_1, n_2, n_3, n_4}}{\Den{4}{n_1, n_2, n_3, n_4}}.
\label{eq:Ev4PC}
\end{eqnarray}
Here $\Num{2}{n_1,n_2}$ and $\Den{2}{n_1,n_2}$ could be obtained as:
\begin{subequations}
\begin{eqnarray}
&&\Num{2}{n_1,n_2}=Q_{n_1,1} Q_{n_2,1}-Q_{n_1+n_2,2}, \\
&&\Den{2}{n_1,n_2}=\Num{2}{0,0}  = Q_{0,1}^2-Q_{0,2}\,. \qquad \qquad \qquad \qquad
\label{eq:2pCorrelation}
\end{eqnarray}
\end{subequations}
Similarly, one can calculate ${\rm N}\langle 4 \rangle_{n_1,n_2,n_3,n_4}$ and  ${\rm D}\langle 4 \rangle_{n_1,n_2,n_3,n_4}$ as follows:
\begin{subequations}
\begin{eqnarray}
&&{\rm N}\langle4\rangle_{n_1, n_2, n_3, n_4}=  Q_{n_1,1} Q_{n_2,1} Q_{n_3,1} Q_{n_4,1}-Q_{n_1+n_2,2} Q_{n_3,1} Q_{n_4,1}
-Q_{n_2,1} Q_{n_1+n_3,2} Q_{n_4,1}\nonumber\\
&&\quad -Q_{n_1,1} Q_{n_2+n_3,2} Q_{n_4,1}+2 Q_{n_1+n_2+n_3,3} Q_{n_4,1}
-Q_{n_2,1}Q_{n_3,1} Q_{n_1+n_4,2}+Q_{n_2+n_3,2} Q_{n_1+n_4,2}\nonumber\\
&&\quad  -Q_{n_1,1} Q_{n_3,1} Q_{n_2+n_4,2}+Q_{n_1+n_3,2} Q_{n_2+n_4,2}+2 Q_{n_3,1} Q_{n_1+n_2+n_4,3}
-Q_{n_1,1} Q_{n_2,1} Q_{n_3+n_4,2}\nonumber\\
&&\quad  +Q_{n_1+n_2,2}Q_{n_3+n_4,2} +2 Q_{n_2,1} Q_{n_1+n_3+n_4,3}+2 Q_{n_1,1} Q_{n_2+n_3+n_4,3}
-6 Q_{n_1+n_2+n_3+n_4,4}\,,\\
&&{\rm D}\langle4\rangle_{n_1, n_2, n_3, n_4}=\Num{4}{0,0,0,0}
=Q_{0, 1}^4 - 6 Q_{0, 1}^2 Q_{0, 2} + 3 Q_{0, 2}^2
+ 8 Q_{0, 1} Q_{0, 3} - 6 Q_{0, 4}\,.
\label{eq:4pCorrelation}
\end{eqnarray}
\end{subequations}
The analogous results for higher-order correlators and differential flow can be written out in a similar manner. The details can be found in~\cite{Bilandzic:2013kga}.

Last but not least, the generic framework not only correct the NUA and NUE effects exactly and efficiently, it can also be applied in  any order of multi-particle correlations for the cases where their direct implementation was not feasible before. For instance, Eqs.~(\ref{eq:Ev2PC}) and~(\ref{eq:Ev4PC}) could be used in Symmetric cumulants $SC(4,2)$ (discussed in Sec.V) by calculating 4-particle correlation of $\langle 4 \rangle_{4,2,-4,-2}$, and 2-particle correlations $\langle 2 \rangle_{2,-2}$ and $\langle 2 \rangle_{4,-4}$.

\section{Extracting the QGP viscosity from flow harmonics}\label{sec:artwork}
\subsection{Semi-quantitative extractions of the QGP shear viscosity}
\quad The hydrodynamic calculations from different groups have shown that the flow harmonics are sensitive to the QGP shear viscosity $\eta/s$, which can be used to study the transport properties of the hot QCD matter~\cite{Heinz:2013th,Gale:2013da,Teaney:2009qa,Song:2013gia,
Romatschke:2009im,Huovinen:2013wma,Romatschke:2007mq,Luzum:2008cw,Song:2007fn,Song:2007ux,Song:2009gc,Dusling:2007gi,Molnar:2008xj,Bozek:2009dw,Chaudhuri:2009hj,
Schenke:2010rr}. Around 2008, the INT group made an early extraction of the QGP shear viscosity
from the integrated and differential elliptic flow data in 200 A GeV Au--Au collisions, using
the 2+1-d viscous simulations with optical Glauber and KLN initializations~\cite{Romatschke:2007mq,Luzum:2008cw}.  They found these two initial conditions  bring large uncertainties for the extracted value of $\eta/s$ around O(100\%).
However, it is not reliable to directly read the  value of $\eta/s$ from a direct model to data comparison since
their model calculation neglect the high viscous and even off equilibrium hadronic evolution,
which only treat such stage as a pure viscous fluid expansion with both chemical and thermal equilibrium. Ref~\cite{Luzum:2008cw,Song:2008hj} further estimated the effects from the late hadronic evolution and
concluded that the extracted value of the specific QGP shear viscosity $(\eta/s)_{QGP}$ can not
exceed an upper limit around $5\times\frac{1}{4\pi}$.

\begin{figure*}[t]
\center
  \includegraphics[width=0.9\linewidth, height=5.8cm]{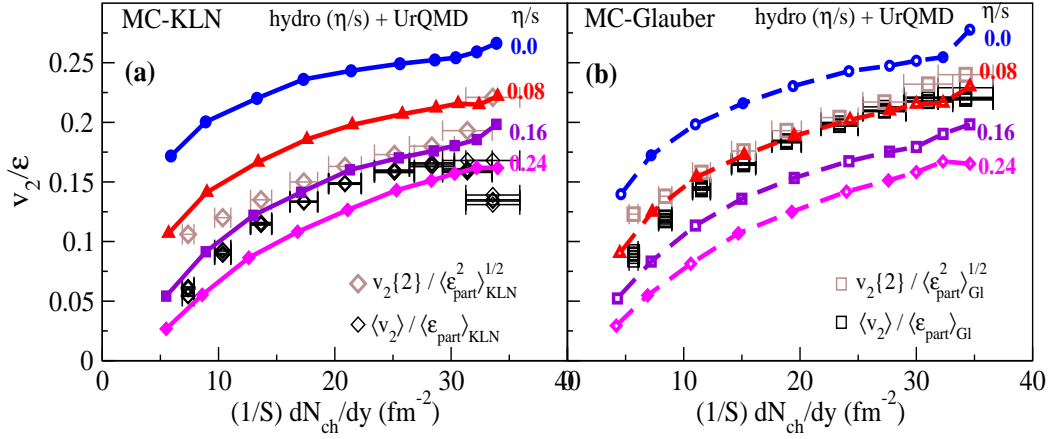}
\caption{(Color online) Eccentricity-scaled elliptic flow as a function of final multiplicity per
 area. The theoretical results are calculated from VISHNU hybrid model calculations with MC-Glauber (left) and MC-KLN (right) initial conditions~\cite{Song:2010mg}. The experimental data are taken from Ref.~\cite{Ollitrault:2009ie}.}
\end{figure*}

For a realistic description for the evolution and decoupling of the hadronic matter, the OSU-LBL group further developed the VISHNU hybrid model~\cite{Song:2010aq} that combines the 2+1-d viscous hydrodynamics with a hadron cascade model-{\tt UrQMD},
and then made a semi-qualitative extraction of the QGP shear viscosity from the integrated elliptic flow data in 200 A GeV Au--Au collisions~\cite{Song:2010mg,Song:2011hk}. Fig.~1 shows the eccentricity-scaled integrated elliptic flow, calculated from VISHNU with MC-Glauber and MC-KLN initial conditions together with a comparison with the corrected experimental data with the non-flow and fluctuation effects removed~\cite{Ollitrault:2009ie}. From Fig.~1, one finds $\frac{1}{4\pi}<(\eta/s)_{QGP}< 2.5\times\frac{1}{4\pi}$, where the main uncertainties of the extracted $(\eta/s)_{QGP}$ are still come from the undetermined initial conditions.  Meanwhile, the corresponding VISHNU simulations with both MC-Glauber and MC-KLN initial conditions could nicely describe the $p_T$-spectra and differential elliptic flow harmonics $v_2(p_T)$  for all charged and identified hadrons at various centrality bins in 200 A GeV Au--Au collisions~\cite{Song:2011hk}. Compared with the early extractions in Ref.~\cite{Romatschke:2007mq}, the precision of the extracted value of $(\eta/s)_{QGP}$ is largely increased due to a better description of the highly viscous hadronic stage.

In Ref.~\cite{Song:2011qa}, the VISHNU simulations were further extrapolated to the LHC energies, which systematically investigated the soft hadron data in 2.76 A TeV the Pb--Pb collisions. The related calculations have showed, with the same $(\eta/s)_{QGP}$ extracted at top RHIC energies, VISHNU slightly over-predicts the ALICE flow data at the LHC.  After slightly increasing $(\eta/s)_{QGP}$ (for the MC-KLN initial conditions, $(\eta/s)_{QGP}$ increases from $\sim$0.16 to $\sim$ 0.20), VISHNU achieves a better description of the elliptic flow of all charged hadrons at varies centralities~\cite{Song:2011qa}.
\begin{figure*}[tb]
   \begin{center}
     \includegraphics[width=7.5cm,height=5.8cm]{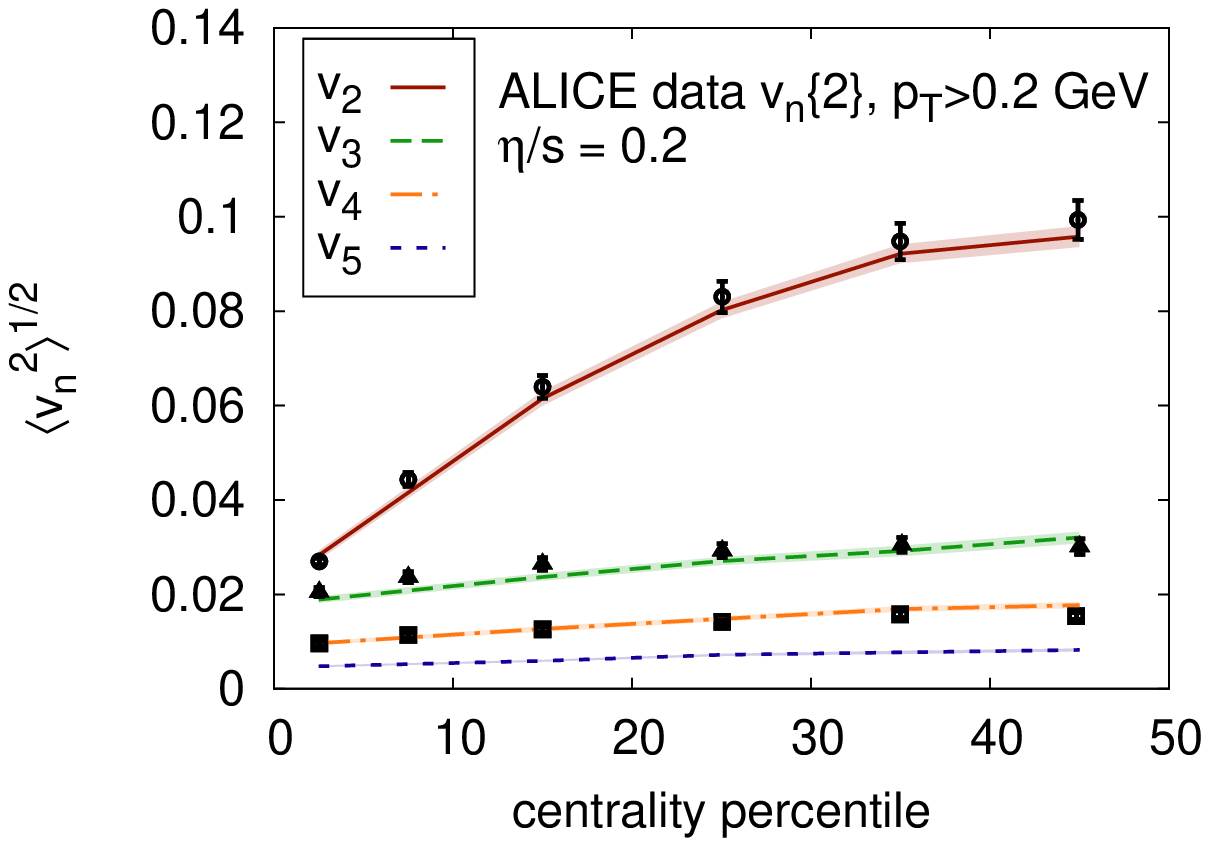}
     \includegraphics[width=7.5cm,height=5.8cm]{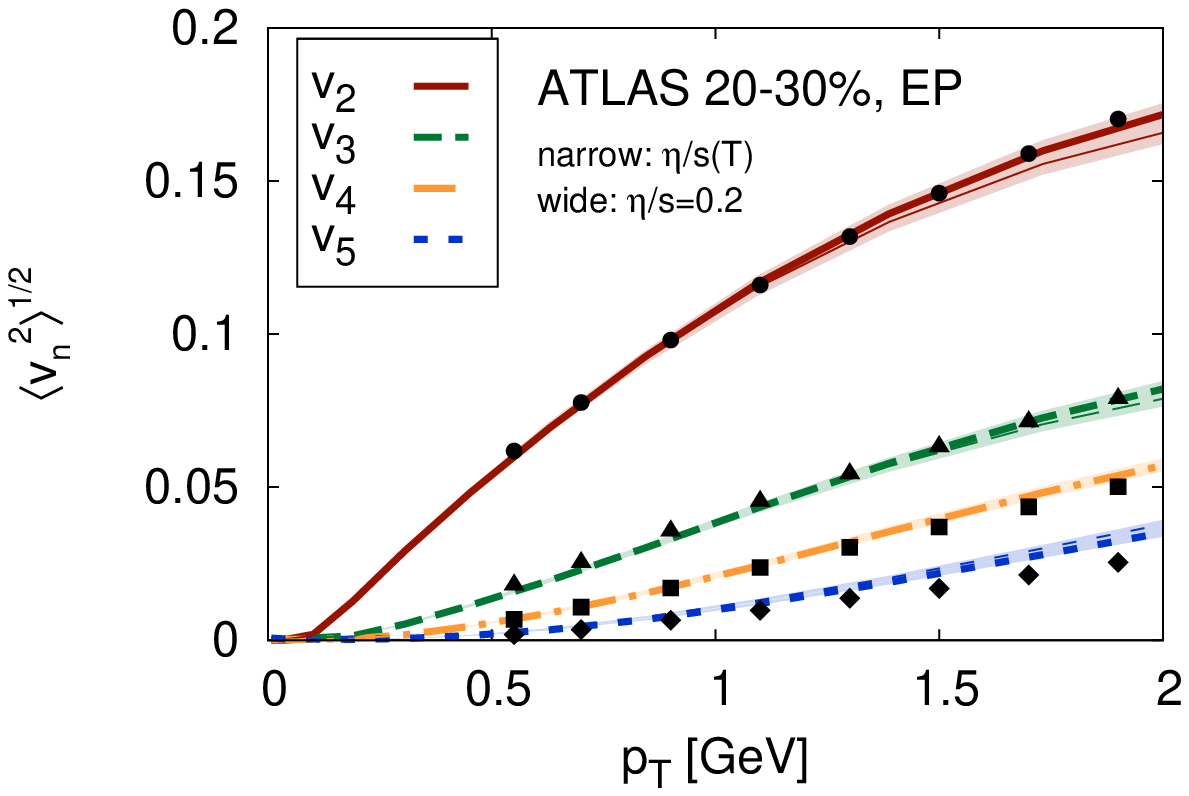}
     \vspace{0.0cm}
     \caption{(Color online) Root-mean-square anisotropic flow coefficients $\langle v_n^2 \rangle ^{1/2}$ and
     $v_n(p_T)$ in 2.76 A TeV Pb--Pb collisions. The theoretical curves are calculated from
     MUSIC with IP-Glasma initial conditions~\cite{Gale:2012rq}. The experimental data in left and right panels are measured by the ALICE collaboration \cite{ALICE:2011ab} and the ATLAS collaboration, respectively.}
     \label{fig:vnCent}
   \end{center}
   \vspace{-0.5cm}
\end{figure*}

\begin{figure*}[htb]
	\begin{centering}
		\includegraphics[scale=0.82]{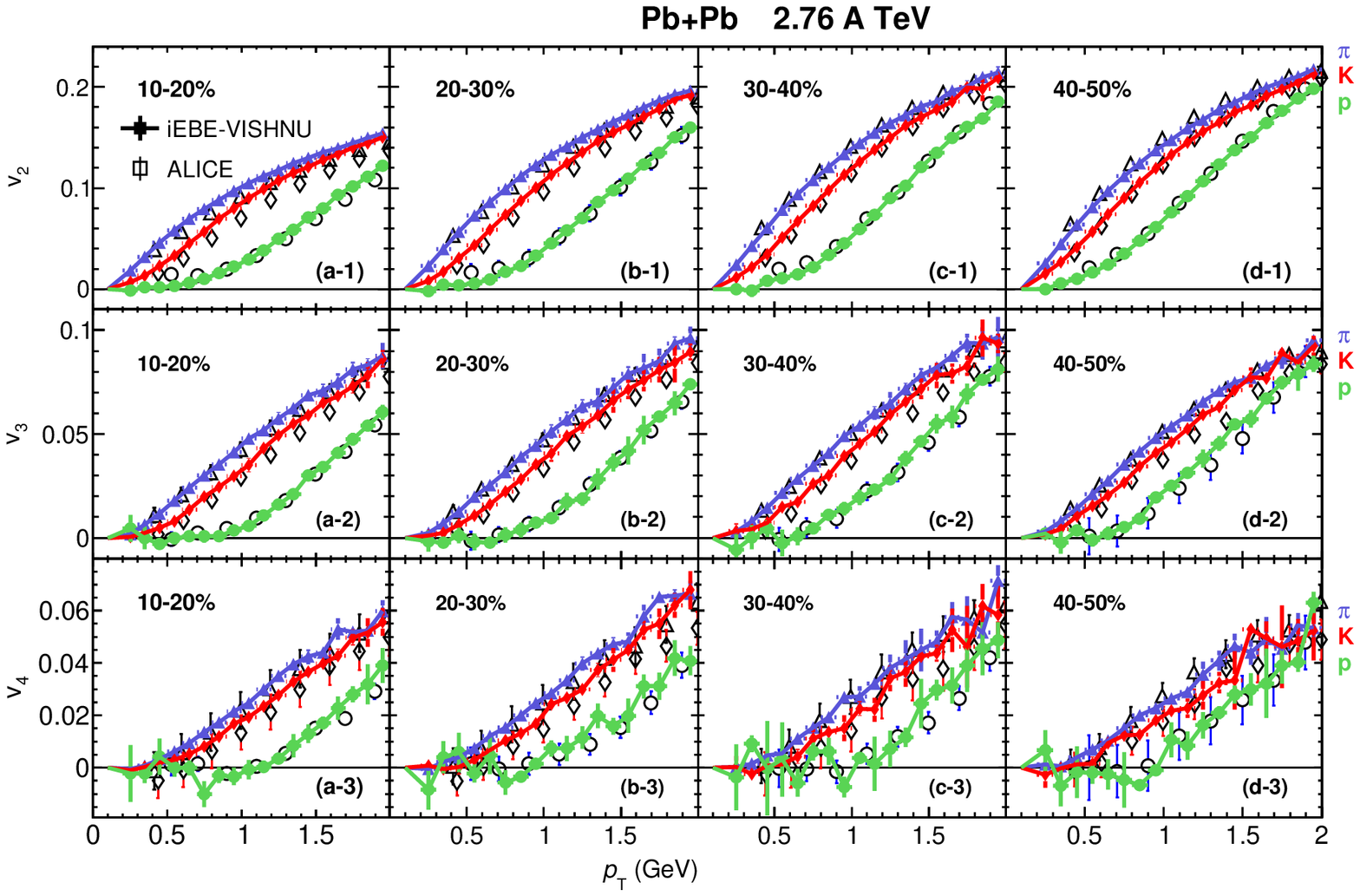}
	\end{centering}
	\vspace{-7mm}
	\caption{(Color online) $v_{n}(p_T)$ ($n=2,3,4$) of pions, kaons and protons in 2.76 A TeV Pb--Pb collisions, calculated from {\tt{iEBE-VISHNU}} with {\tt{AMPT}} initial conditions~\cite{Xu:2016hmp}. The experimental data are taken from the ALICE paper~\cite{Abelev:2014pua,Mohammadi:2016umt}.
	\label{fig:flowharmonics} }
\end{figure*}

Many of the early hydrodynamic or hybrid model simulations (includes these 2+1-d hydrodynamic and VISHNU calculations mentioned above)~\cite{Song:2010aq,Song:2010mg,Song:2011hk,Song:2011qa,Song:2013qma,Heinz:2011kt,Zhu:2015dfa} are belong to the category of single-shot simulations, which input smooth initial energy/entropy profiles from early initial condition models or input some smoothed profiles obtained from averaging  millions of events from some specific fluctuating initial condition models. Correspondingly, the effects from initial state fluctuations are neglected.  Around 2012, the Mcgill group further developed event-by-event 3+1-d viscous hydrodynamic simulations with the IP-Glasma pre-equilibrium dynamics (MUSIC + IP-Glasma) and calculated the flow harmonics at different orders at RHIC and the LHC~\cite{Gale:2012rq}. Fig.~2 shows the integrated and differential $v_n$($n=$ 2...5) of all charged hadrons
in 2.76 A TeV Pb--Pb collisions. Impressively, these different flow harmonic data are nicely described by the MUSIC simulations with $\eta/s=0.2$ or a temperature dependent $\eta/s(T)$ at various centralities. Meanwhile, their simulations also shows the averaged QGP viscosity are slightly larger at the LHC than at RHIC as found in~\cite{Song:2011qa}. Compared with the VISHNU simulations~\cite{Song:2010mg,Song:2011hk, Song:2011qa}, these MUSIC calculations are pure hydrodynamic, which does not specially treat the hadronic evolution with a hadronic afterburner. However, the main results will not be significantly changed since the flow harmonics at the LHC energies are mainly developed (or even reach saturation) in the QGP phase.

For the hydrodynamic simulation with IP-Glasma initial conditions, a balanced initial eccentricities at different order are generated at the beginning, which helps to achieve a simultaneous fit of the elliptic flow, triangular flow and other higher order harmonics. In contrast, the hydrodynamic calculations with either Mc-Glauber or Mc-KLN initial conditions fail to simultaneously describe all the flow harmonics $v_n$ at different order ($n=$ 2 ... 5) although they can nicely fit the elliptic flow data with a well-tuned QGP shear viscosity. Therefore, these higher-order flow harmonic measurements disfavor these two initial conditions,  which also motivated the later developments of other initial condition models. In short, the extracted value of the QGP viscosity may largely influenced by the initial conditions used in the hydrodynamic calculations. Meanwhile, higher order flow harmonics as well as other flow observables (please also refer to Sec. V for details) could put straight constrains for the initial condition models and for the extracted value of the QGP shear viscosity.

Besides these flow data of all charged hadrons, the flow harmonics of identified hadrons could reveal more information on the hadronic evolution of the hot QCD matter and provide additional test for extracted values of the QGP transport coefficients obtained from the soft hadron data of all charged hadrons.   Ref.~\cite{Song:2013qma} and ~\cite{Heinz:2011kt} have shown, for the extracted constant QGP shear viscosity obtained from the elliptic flow in 2.76 A TeV Pb--Pb collisions,  VISHNU hybrid model could nicely describe the differential elliptic flow data of pions, kaons and protons~\cite{Song:2013qma,Heinz:2011kt}. Meanwhile, it could also roughly fit the elliptic flow data of strange and multi-strange hadrons ($\Lambda$, $\Xi$ and $\Omega$) measured at the LHC ~\cite{Zhu:2015dfa}. Recently, the ALICE collaboration further measured the higher order flow harmonics of identified hadrons in 2.76 A TeV Pb--Pb collisions, which showed that the triangular and quadratic flow harmonics of pions, kaons and protons present similar mass ordering as the case for the elliptic flow~\cite{Adam:2016nfo}. In Ref.~\cite{Xu:2016hmp},
the PKU group implement the iEBE-VISHNU hybrid model with the AMPT initial conditions to investigate the flow harmonics of identified hadrons $v_n(p_T)$ ($n=$ 2,3,4) at the LHC.  After tuning the Guassian smearing factor for initial energy decompositions and the QGP shear viscosity, the differential $v_n$ of all charged hadrons can be nicely described by the
iEBE-VISHNU simulations. As show in Fig.~3, iEBE-VISHNU also nicely describes the  $v_n$ data of pions, kaons and protons, especially reproduces correct mass-orderings for these different flow harmonics.
Ref~\cite{Xu:2016hmp} also showed the pure hydrodynamic simulations do not generate enough mass-splittings between the $v_n$ of pions and protons. The late hadronic evolution in the iEBE-VISHNU
re-distributes the anisotropy flow to various hadron species through the microscopic hadronic scatterings, which enhances
the $v_n$ mass splitting between pions and protons and leads to a nice description of the
experimental data~\cite{Xu:2016hmp}.

\vspace{0.15cm}
\underline{\emph{The issues of bulk viscosity}}
\vspace{0.10cm}

For simplicity, the early semi-quantitative extraction of the QGP shear viscosity at RHIC and the
LHC neglects the effects from bulk viscosity~\footnote{At the LHC and top RHIC energies, the heat
conductivity can be neglected due to the almost vanishing net baryon density.}.
The (0+1)-d viscous hydrodynamic calculations without transverse expansion
~\cite{Torrieri:2008ip,Rajagopal:2009yw} suggested that, for a uniform system
undergoing rapid boost-invariant longitudinal expansion, the bulk pressures can turn
into negative values, leading to mechanically unstable fluid with negative effective
pressure. The 2+1-d viscous hydrodynamics with single shoot simulations showed that the bulk viscosity also suppresses the elliptic flow as the shear viscosity~\cite{Song:2009rh,Song:2009je,Monnai:2009ad,Denicol:2009am,
Noronha-Hostler:2014dqa},
but with smaller efforts due to the critical slowing down near the QCD phase transition~\cite{Song:2009rh}.
Recently, the 3+1-d event-by-event simulations from MUSIC found that the bulk viscosity largely influence the average transverse momentum of identified hadrons~\cite{Ryu:2015vwa}. For the MUSIC calculation with the IP-Glasma initial condition, the fitting of the $p_T$ spectra are largely improved by a properly chosen bulk viscosity, which also leads to
a consistent descriptions of other soft hadron data, such as the integrated and differential flow harmonics at various centralities in 2.76 A TeV Pb--Pb collisions.

\begin{figure}[tbh]
    \vspace{5mm}
	\begin{centering}
		\includegraphics[scale=0.9]{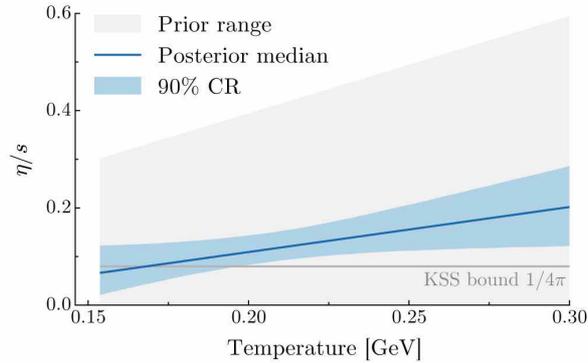}
	\end{centering}
	\vspace{0mm}
	\caption{(Color online) Estimated temperature dependence of the shear viscosity $(\eta/s)(T)$ above the QCD phase transition (for $T_c > 154$ MeV), obtained from a multi-parameter model to data comparison~\cite{Bernhard:2016tnd}.
	\label{fig:flowharmonics} }
\end{figure}

\subsection{Quantitative extractions of the QGP shear and bulk viscosity with massive data evaluations}

For the flow calculations and predictions at RHIC or at the LHC, most of the hydrodynamics or hybrid model simulations, with different type of initial conditions, input a constant value of the specific QGP shear viscosity and neglect the effects
of bulk viscosity. The early model calculations also revealed that the averaged QGP shear viscosity changes with the collision energies, which is slightly larger at the LHC than at RHIC~\cite{Song:2011qa,Gale:2013da,Karpenko:2013ama,Song:2012ua}.
It is thus very important to extract a temperature-dependent QGP shear viscosity $(\eta/s)_{QGP}(T)$ from the massive soft hadron data in relativistic heavy ion collisions. For the purposes of  massive data evaluations,
the Livermore group developed the CHIMERA algorithm (a comprehensive heavy ion model evaluation and reporting algorithm), and  extracted of the initial temperature and the QGP shear viscosity from a simultaneous fit of the $p_T$ spectra, elliptic flow, and HBT radii in 200 A GeV Au + Au collisions~\cite{Soltz:2012rk}. Note that this early massive hydrodynamic simulations around 2012 assume the QGP shear viscosity is a constant value and the bulk viscosity is zero, together with an input of the traditional MC-Glauber initial condition which has been ruled out by some later flow measurements.

To avoid the limitations of simultaneously tuning multiple free parameters in the early work~\cite{Soltz:2012rk}, the Duke-OSU group implemented the Bayesian method to the event-by-event hybrid model simulations~\cite{Bernhard:2015hxa}, and then quantitatively estimated the properties of the QGP through a multi-parameter model to data comparison, using the  parametric  T\raisebox{-.5ex}{R}ENTo initial conditions~\cite{Bernhard:2016tnd}. With the new developed massive data evaluating techniques, the global fitting of the multiplicity, transverse momentum, and flow data at the LHC  constrain the free parameters in the T\raisebox{-.5ex}{R}ENTo model, which also give an extracted temperature-dependent specific shear viscosity and bulk viscosity.

\begin{figure*}[tbh]
	\begin{centering}
		\includegraphics[width=15.0cm,height=6.2cm]{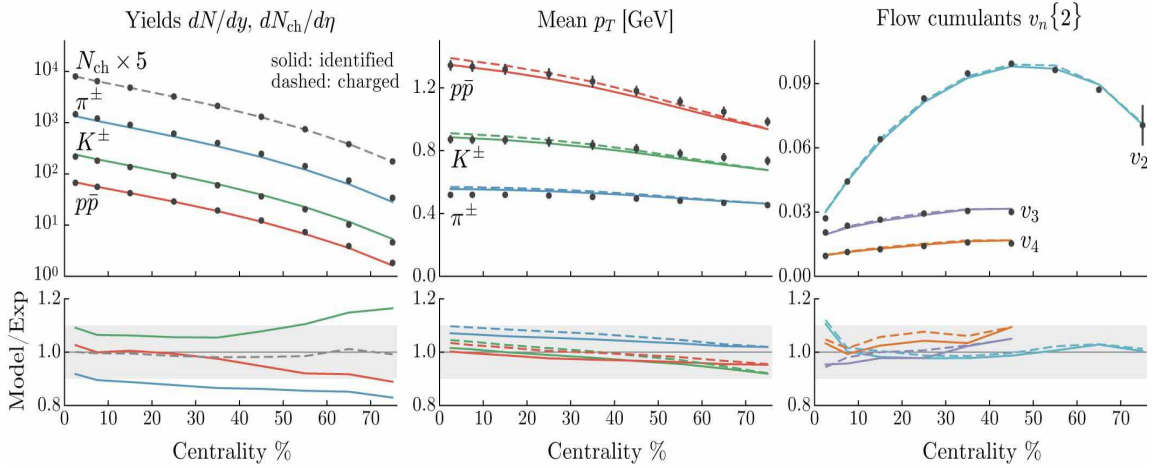}
	\end{centering}
	\vspace{0.0mm}
	\caption{(Color online) Multiplicities, mean $p_T$ of all charged and identified hadrons
and the integrated $v_n$ ($n=$ 2,3,4) of all charged hadrons in 2.76 A TeV Pb--Pb collisions, calculated from
event-by-event hybrid model with the high-probability
parameters extracted from the massive data fitting~\cite{Bernhard:2016tnd}.
The data are from the ALICE experiment~\cite{Abelev:2013vea,ALICE:2011ab}.}
   \vspace{-0.3cm}
\end{figure*}

Fig.~4 shows the estimated temperature dependent shear viscosity $(\eta/s)(T)$ from the DUKE-OSU group, obtained from the massive data fitting in 2.76 A TeV Pb+Pb collisions. The blue line is the median with a blue band showing the 90\% credible region. Correspondingly, a nonzero bulk viscosity with a peak near the QCD phase transition has also been extracted simultaneously (For details, please refer to~\cite{Bernhard:2016tnd}).
With these extracted QGP transport coefficients other extracted most probable parameters, the event-by-event hybrid simulations give an excellent overall fit for the multiplicities and mean $p_T$ of all charged and identified hadrons
and the integrated $v_n$ ($n=$ 2,3,4) of all charged hadrons from the most central collisions to the peripheral collisions in Pb--Pb collisions at the LHC, as shown in Fig.~5.

Note that this extracted  $\eta/s(T)$, within the uncertainty band, is compatible with the well-known KSS bound $\eta/s < 1/4\pi$  \cite{Danielewicz:1984ww, Policastro:2001yc, Kovtun:2004de}, which also supports several early semi-quantitative extractions of the QGP viscosity at RHIC and the LHC.  For example, the extracted specific QGP viscosity  $\frac{1}{4\pi}<(\eta/s)_{QGP}< 2.5\times\frac{1}{4\pi}$ from
the VISHNU calculations with MC-Glauber and MC-KLN initial conditions~\cite{Song:2010mg,Song:2011hk} and the
implemented $\eta/s = 0.095$ (with the same bulk viscosity parametrization ) in the MUSIC simulations with the IP-Glasma initialization~\cite{Ryu:2015vwa} are both consistent with this quantitative extracted results from the DUKE-OSU collaborations. The early EKRT viscous hydrodynamic calculations for the flow data at RHIC and the LHC  also prefer a temperature-dependent $\eta/s(T)$ with a positive slope~\cite{Niemi:2015qia}.

Compared with the early extraction of the QGP viscosity with specific initial condition,  Ref.~\cite{Bernhard:2016tnd} implement the parametric {\tt T\raisebox{-.5ex}{R}ENTo} model that could smoothly interpolates among various
initial condition schemes through tuning the related parameters. It is thus an ideal initial state model for the massive  model-to-data comparison, which helps to make a simultaneous constraint for the initial conditions and the QGP transport
coefficients. It was found that initial entropy deposition from the constrained T\raisebox{-.5ex}{R}ENTo model with fixed parameters is approximately proportional to the geometric mean of the participant nuclear densities, which gives similar scaling as the successful EKRT and IP-Glasma initial conditions.

In Ref.~\cite{Auvinen:2016tgi}, the Bayesian statistical analysis was extended to the massive data fitting in Au--Au collisions at $\sqrt{s_{\rm NN}}=$ 19.6, 39 and 62.4 GeV. It was found that the extracted constant QGP specific shear viscosity $\eta/s$ decreases with the increase of collision energy, which shows a similar result obtained from the early hybrid model calculations~\cite{Karpenko:2013ama}. In the future, a combined massive data fitting at RHIC (including BES) and the LHC will give more precise temperature-dependent transport coefficients of the QGP.

\section{Initial state fluctuations and final state correlations}

The event-by-event initial state fluctuations of the created QGP fireballs lead to the final state correlations, which produce the elliptic flow, triangular flow and other higher-order flow harmonics as observed in the experiments at RHIC and the LHC~\cite{Alver:2010gr,Alver:2010dn,Adare:2011tg,Gardim:2011xv,Adamczyk:2013waa,ALICE:2011ab,Aamodt:2011by,ATLAS:2012at}.
The QGP viscosity largely suppresses flow harmonics at different order $v_n$. As reviewed in last section, the transport properties of the QGP fireball have been extracted from these flow data with the event-by-event hydrodynamics / hybrid model simulations\cite{Song:2011qa, Schenke:2012fw,Gale:2012rq}. In this section, we will review other flow observable, such as event-by-event $v_n$ distribution,  the event plane correlations, the correlations of flow harmonics, etc., that are more sensitive to the details of model calculations, which may provide additional constrains for the initial state models and
for the extracted QGP transport coefficients in the future.
\begin{figure}[tbh]
   \begin{center}
     \includegraphics[width=6.75cm,height=12cm]{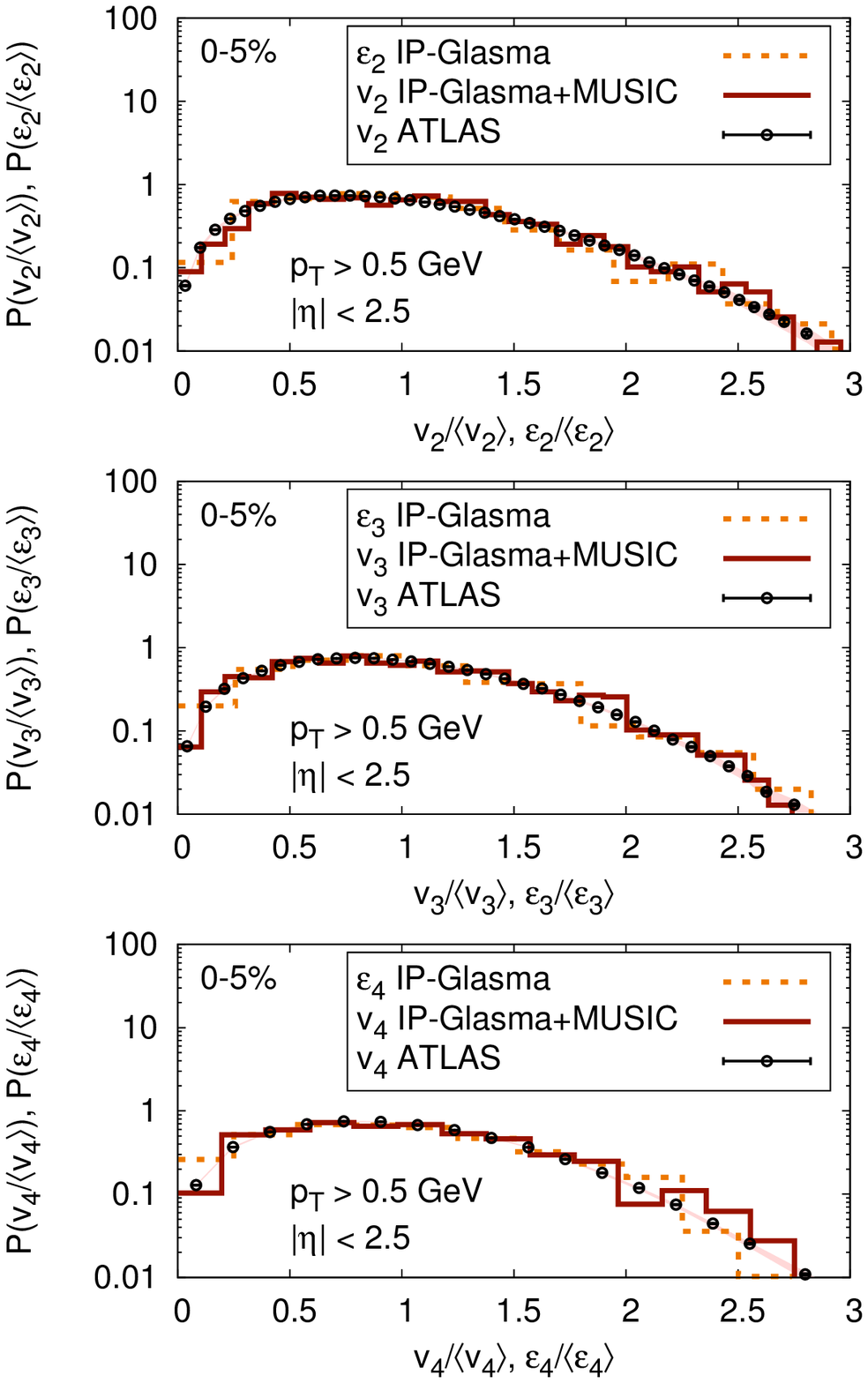}
     \includegraphics[width=6.75cm,height=12cm]{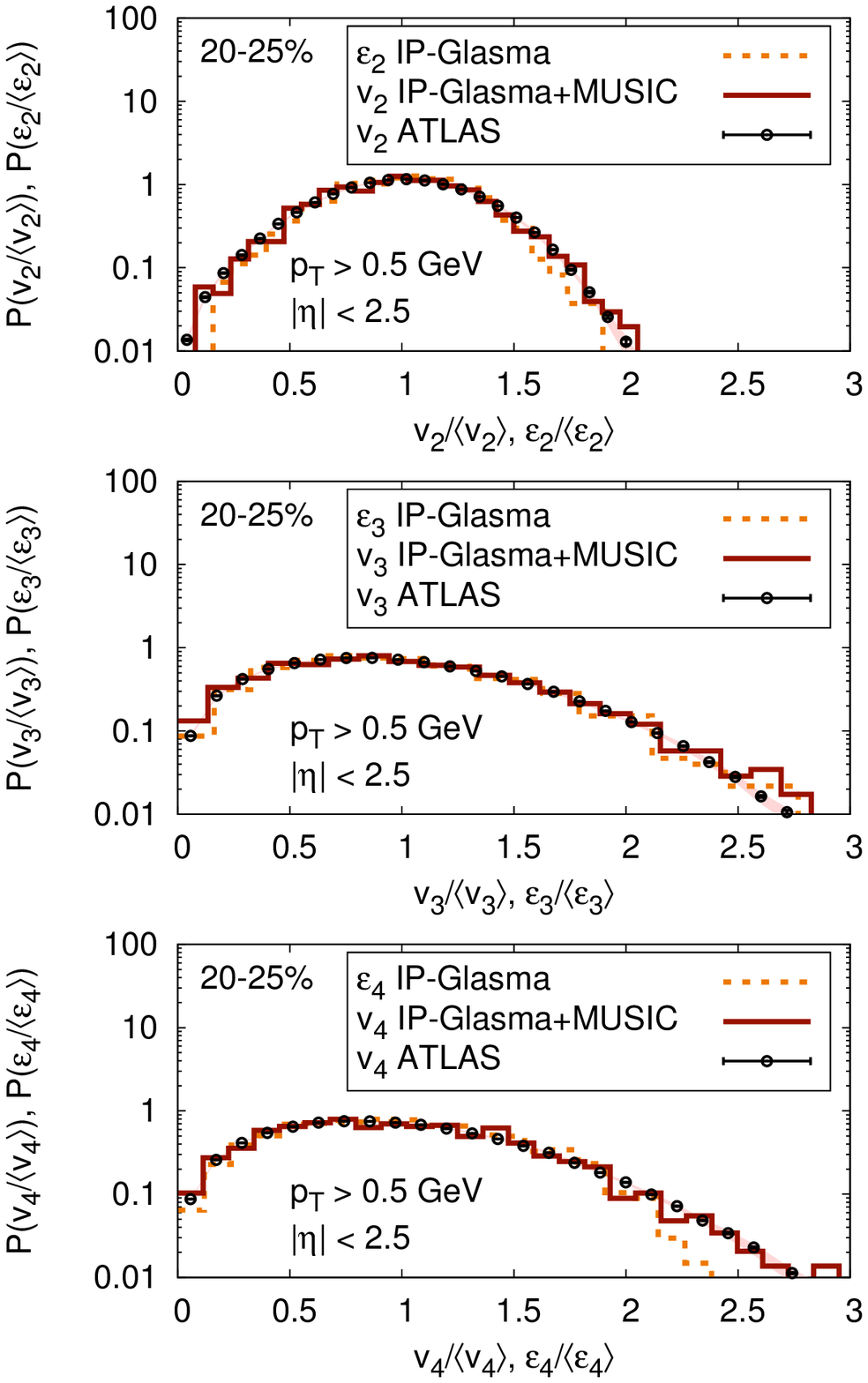}
     \caption{(Color online) Scaled event-by-event distributions of $v_n$ $($n=$ 2,3,4)$ from MUSIC simulations with the IP-Glasma initial conditions~\cite{Gale:2012rq,Gale:2013da}, together with a comparison with the ATLAS data~\cite{Aad:2013xma}.}
     \label{fig:vnenDist-20-25}
   \end{center}
   \vspace{-0.7cm}
\end{figure}

\vspace{0.2cm}
\underline{\emph{Event-by-event $v_n$ distribution}}:
\vspace{0.10cm}

The flow harmonics $v_{n}$ are generally measured within a base of the event-average, which
mainly reflects the hydrodynamic response to the averaged initial eccentricity coefficients
$\varepsilon_{n}$ within some centrality bin.  With a large amount of particles produced per event at the LHC,
a direct measurement of the event-by-event $v_{n}$ distribution becomes possible, which provides
more information on the initial state fluctuation and the underlying probability density function.
Around 2012, the ATLAS Collaboration made the first measurement of the event-by-event distributions of $v_{n}$ ($n = 2,\ 3, \ 4$) in Pb--Pb collisions at $\sqrt{s_{\rm NN}}=$ 2.76 TeV~\cite{Aad:2013xma}. Fig.~6 shows the MUSIC hydrodynamic calculations nicely describe the ATLAS data with the IP-Glamsa initial conditions. It also shows, for n=2 and 3,  the rescaled  $v_{n} / \left< v_{n} \right>$ distributions mostly follow the ${\varepsilon_n}/{\left< \varepsilon_n \right>}$ distributions from the initial state, which are not sensitive to the details of the hydrodynamic evolution~\cite{Gale:2012rq}. Due to the mode couplings effects for higher flow harmonics,
the distributions of $v_{4} / \left< v_{4} \right>$  are not necessarily  follow  ${\varepsilon_4}/{\left< \varepsilon_4 \right>}$, especially for non-central Pb+Pb collisions. The hydrodynamic evolution balances the distributions of $v_{4} / \left< v_{4} \right>$, making a nice description of the experimental data. In Ref.~\cite{Aad:2013xma},  the measured $v_n$ distributions were compared with the $\varepsilon_n$ distributions from MC-Glauber and MC-KLN models, which demonstrated certain deviations between model and data for most of the centrality classes. The $v_n$ distributions thus provide strong constrains on the initial state models, which do not favor the MC-Glauber and MC-KLN initial conditions.
\begin{figure*}[tbh]
\centering
\includegraphics[width=15.0cm,height=5.0cm]{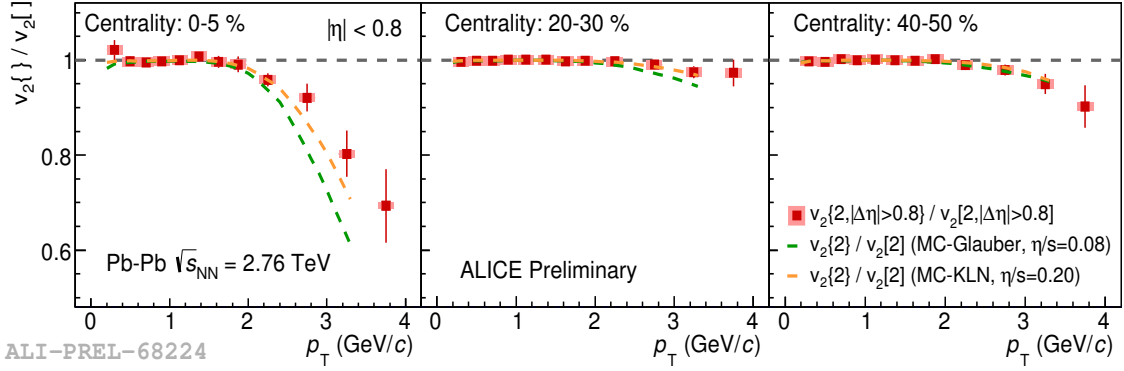}
\caption{(Color online) The ratio $v_{n}\{2\}/v_{n}[2]$ at various centralities in 2.76 A TeV Pb--Pb collisions. The theoretical lines are calculated from VISH2+1 with MC-Glauber and MC-KLN initial conditions~\cite{Heinz:2013bua}, the experimental data are measured by the ALICE collaborations~\cite{Zhou:2014bba}.}
\label{fig:ptdepV}
\end{figure*}

The ATLAS measurements can also be used to examining the underlying $p.d.f.$  of the $v_{n}$ distributions. The most popular parameterizations are the Bessel-Gaussian distributions~\cite{Voloshin:2007pc}:
\begin{equation}
p(v_{n}) = \frac{v_{n}}{\sigma^{2}} I_{0}\left(  \frac{v_{n} v_{n}}{\sigma^{2}} \right) {\rm exp} \left( - \frac{ v_{0}^{2} + v_{n}^{2}} {2\sigma^{2}}  \right) ,
\end{equation}
where $v_{0}$ is the anisotropic flow from the reaction plane $\Psi_{\rm RP}$ and $\sigma$ is the anisotropic flow fluctuation.
It was reported that the Bessel-Gaussian distribution could nicely describe the $v_{2}$ distributions for mid-central collisions~\cite{Voloshin:2007pc, Broniowski:2007ft}. Without the constraint of $\varepsilon_{2} <$ 1 for each event, it is not expected to work well in peripheral collisions~\cite{Yan:2014afa}.
To fix this problem, a new function, named ``Elliptic Power" distribution, was proposed in~\cite{Yan:2014afa}, which are expressed as:
\begin{equation}
p(v_{n}) = \frac{\alpha \, v_{n}}{\pi}  \left( 1- v_{0}^{2} \right)^{\alpha + \frac{1}{2}}  \int_0^{2\pi} \frac{ \left( 1- v_{n}^{2} \right)^{\alpha - 1} \, \mathrm{d}\phi} { \left(  1- v_{0} \,v_{n} \cos \phi  \right)^{2\alpha +1}},
\end{equation}
where $\alpha$ quantifies the fluctuations and $v_{0}$ has the same meaning as the Bessel-Gaussian parameterizations.
As a promising candidate of underlying $p.d.f.$ of $v_{n}$ distribution, the Elliptic-Power function can nicely describe the event-by-event $v_{2}$ and $v_{3}$ distributions~\cite{Yan:2014afa,Zhou:2015eya}. However, it can not give an equally nice fitting for these distributions of higher flow harmonics ($n\geqslant 4$), which are largely influenced by the non-linear hydrodynamic response. For details, please refer to~\cite{Yan:2014afa,Zhou:2015eya}.

\vspace{0.2cm}
\underline{\emph{De-correlations of the flow-vector $V_{n}$}}:
\vspace{0.10cm}

Recently, it was realized that the produced particles at different transverse momentum $p_T$ and rapidity y do not share a common flow angle or event plane. Such transverse momentum and rapidity dependent flow angles fluctuate event-by-event, which
also breaks the factorizations of the flow harmonics~\cite{Heinz:2013bua,Gardim:2012im}. To evaluate the de-correlations of the flow-vector, especially on the transverse momentum dependence, two new observables, $v_{n}\{2\}/v_{n}[2]$ and the factorization ratio $r_{n}$, have been proposed, which are defined as:
\begin{eqnarray}
\frac{v_{n}\{2\}}{v_{n}[2]}(p_{\rm T}^{a}) &=& \frac{ \left< v_{n}^{a} v_{n} \cos \left[ n \left( \Psi_{n}^{a} - \Psi_{n} \right) \right]  \right>}  {\left< v_{n}^{a} v_{n}^{a} \right>^{1/2} \left< v_{n} v_{n} \right>^{1/2}  }; \\
r_{n} &=& \frac{ \left< v_{n}^{a} v_{n}^{b} \cos \left[ n \left( \Psi_{n}^{a} - \Psi_{n}^{b} \right) \right]  \right>}  { \left< v_{n}^{a} v_{n}^{a} \right>^{1/2}  \left< v_{n}^{b} v_{n}^{b} \right>^{1/2}  }
\end{eqnarray}
where $v_{n}^{a}$, $\Psi_{n}^{a}$ (or $v_{n}^{b}$, $\Psi_{n}^{b}$) are the $n^{th}$-order flow harmonics and flow angle at the transverse momentum $p_{\rm T}^{a}$ (or $p_{\rm T}^{b}$).
The $p_{\rm T}$ dependent fluctuations of the flow angle and magnitude make $v_{n}\{2\}/v_{n}[2] < $ and $r_n$ deviated from 1. As shown in Fig.~\ref{fig:ptdepV}, these deviations from unity have already been observed in experiment and qualitatively described by the related hydrodynamic calculations~\cite{Heinz:2013bua}, which indicates the existence of the $p_{\rm T}$ dependent fluctuations of flow angle and magnitude.
\begin{figure*}[tbh]
\centering
 \includegraphics[width=0.95\linewidth, height=7cm]{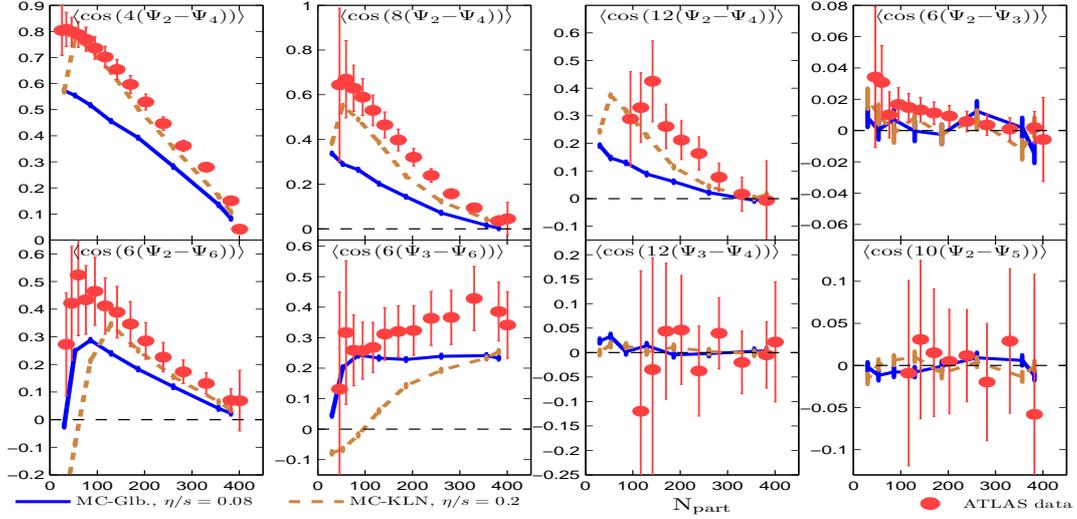}
  \caption{(Color online) Centrality dependent event-plane correlations, calculated from event-by-event $\tt{VISH2+1}$ hydrodynamic simulations with MC-Glauber and MC-KLN initial conditions~\cite{Qiu:2012uy}. The data are measured by the
  ATLAS collaborations~\cite{Aad:2014fla}.}
  \label{fig:heinz2}
\end{figure*}

The fluctuations in the longitudinal direction have also been investigated both in experiment and in theory~\cite{Khachatryan:2015oea,Pang:2014pxa,Pang:2015zrq,Xiao:2015dma,Ma:2016fve}. Ref.~\cite{Pang:2015zrq} found that the
the final state de-correlations of the anisotropic flows in different pseudo rapidity regime is associated with the
spatial longitudinal de-correlations from the initial state. It also predicted a larger longitudinal
decorrelations at RHIC than the ones at the LHC, which provide opportunities to further study
the longitudinal fluctuation structures of the initial stage.

\vspace{0.2cm}
\underline{\emph{Event-plane correlations}}:
\vspace{0.10cm}

The correlations  between different flow vectors  could reveal more information on the initial state fluctuations and the hydrodynamic response~\cite{Zhou:2016eiz}.
In Ref.~\cite{Aad:2014fla}, the ATLAS Collaboration has measured the event-plane correlations
among two or three event-plane angles, $\langle \cos(c_nn\Psi_n + c_mm\Psi_m) \rangle$ and $\langle \cos(c_nn\Psi_n + c_mm\Psi_m + c_hh\Psi_h) \rangle$, in 2.76 A TeV Pb--Pb collisions and observed several different centrality-dependent trends for these correlators. It was also reported that the MC-Glauber model, which only involves the correlations from the initial state, can not reproduce the trends for many of these correlators~\cite{Aad:2014fla}.  Using event-by-event hydrodynamics with MC-Glauber and MC-KLN initial conditions, Qiu and Heinz have systematically calculated the event-plane correlations and demonstrated the hydrodynamic evolution is essential for an overall qualitative description of various flow angle correlations~\cite{Qiu:2012uy}. Fig.~8 presents the model to data comparisons for several selected correlations functions which shows, although correlation strength is sensitive to the initial conditions and the QGP shear viscosity, hydrodynamics  successfully reproduce the centrality-dependent trend of these event-plane correlations. In contrast, the correlations of the initial eccentricity plane show large discrepancies with the measured and calculated event correlations of the final produced particles, including magnitudes, qualitative centrality dependence, and even in signs~\cite{Qiu:2012uy}. In~\cite{Aad:2014fla, Bhalerao:2013ina},  it was found that the AMPT simulations are also able to roughly reproduce the ATLAS data with well tuned parameters. These different model calculations involving final state interactions~\cite{Qiu:2012uy,Aad:2014fla,Bhalerao:2013ina} demonstrate that the observed event-plane correlations are not solely driven by the initial geometry, but largely influenced by the complicated evolution of the QGP fireball.

\begin{figure*}[t]
\centering
\includegraphics[width=0.49\textwidth]{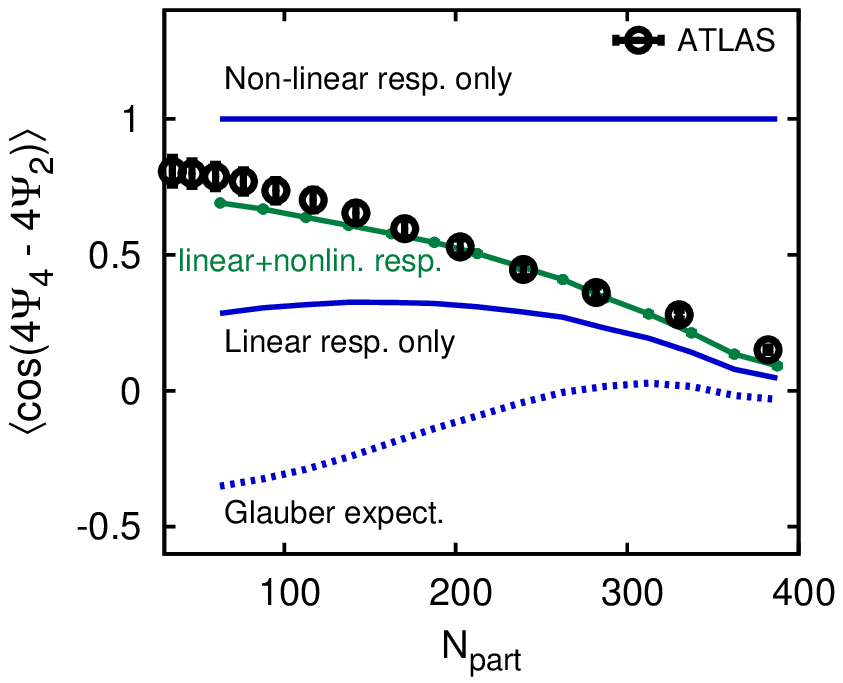}
\includegraphics[width=0.49\textwidth]{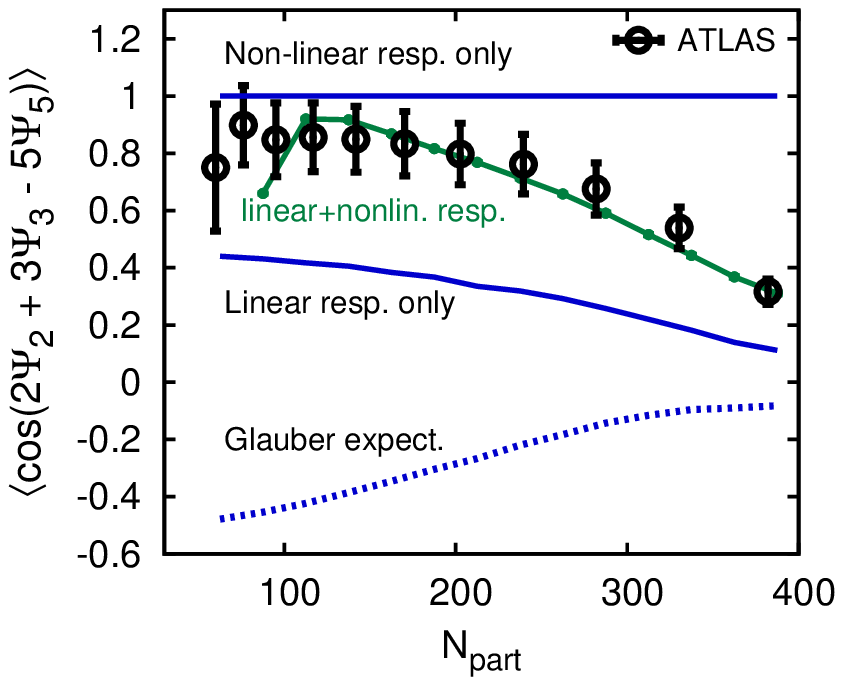}
\caption{(Color online) The separate contributions from linear, non-linear and combined response to the event-plane correlations~\cite{Teaney:2013dta}, together with a comparison with the ATLAS data~\cite{Aad:2014fla}.}
\label{fig:yan}
\end{figure*}

Using a nonlinear response formalism, Ref.~\cite{Teaney:2013dta} calculated the event plane correlations from the initial energy density expanded with the cumulants method,which roughly reproduces the centrality-dependent trends of several selected correlations. It is also found that the non-linear response of the medium have strong influence on these related correlators. As shown in Fig.~9, the linear response alone is not able to describe the $\langle \cos(4(\Psi_2 - \Psi_4)) \rangle$ and $\langle \cos(2\Psi_2 + 3\Psi_3 - 5\Psi_5) \rangle$ correlators,  while a good description of the data can be achieved after combining the contributions of both linear and non-linear response.


\begin{figure*}[tbh]
\centering
\includegraphics[width=0.95\textwidth]{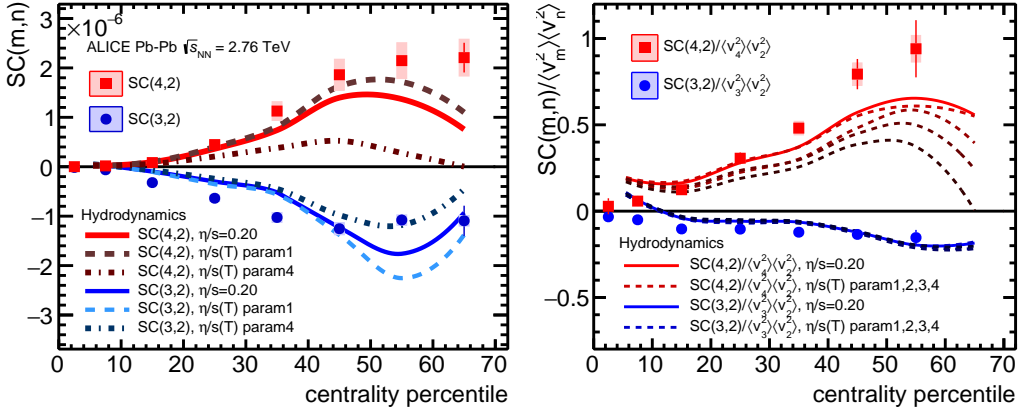}
\caption{(Color online) The centrality dependence of symmetric cumulants $SC(4,2)$ and $SC(3,2)$ in 2.76 A TeV Pb--Pb collisions~\cite{ALICE:2016kpq}. }
\label{fig:sc_ALICE}
\end{figure*}

\begin{figure*}[tbh]
\centering
\includegraphics[width=0.95\linewidth,height=10.0cm]{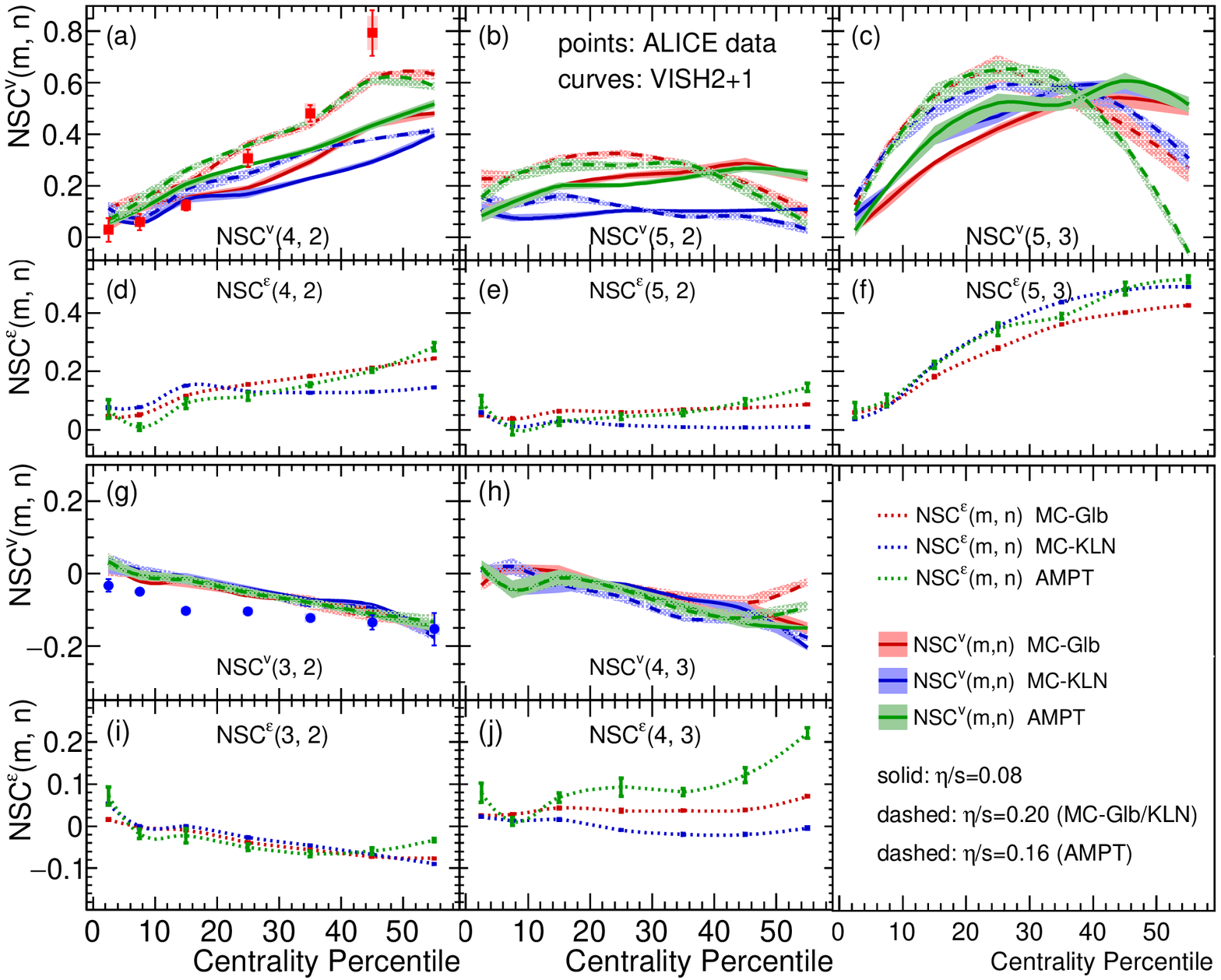}
\caption{(Color online) The centrality dependence of normalized symmetric cumulants ${\rm NSC}(m,n)$ and the corresponding normalized symmetric cumulants of the initial eccentricity coefficients ${\rm NSC}^{\varepsilon}(m,n)$ in 2.76 A TeV Pb--Pb collisions, calculated from event-by-event VISH2+1 simulations with MC-Glauber, MC-KLN and AMPT initial conditions. ~\cite{Zhu:2016puf}.}
\label{fig:sc_hydro4}
\end{figure*}

\vspace{0.2cm}
\underline{\emph{Correlations of flow harmonics}}:
\vspace{0.10cm}

Besides the event-plane correlations, the correlations between different flow harmonics are other important observables closely related to the corrections of the flow vectors, that could further reveal the initial state correlations and the hydrodynamic response. Using the Event-Shape Engineering (ESE)~\cite{Schukraft:2012ah}, the ATLAS Collaboration firstly measured the correlations between flow harmonics based on the 2-particle correlations
and found that $v_{2}$ and $v_{3}$ are anti-correlated, $v_{2}$ and $v_{4}$ are correlated~\cite{Aad:2015lwa}~\footnote{For the related qualitative investigations from hydrodynamics, please refer to~\cite{Qian:2016pau}}. Recently, a new observable, called Symmetric Cumulants $SC^{v}(m, n)$, were proposed as an alternative approach to measure the correlations between different flow harmonics. It is defined as $SC^{v}(m, n)= \left< v_{m}^{2} \, v_{n}^{2} \right> - \left< v_{m}^{2} \right> \left< v_{n}^{2} \right>$ and can be measured by the multi-particle cumulant method.  The related Monte-Carlo model simulations imply that $SC^{v}(m, n)$ is insensitive to the non-flow effects~\cite{ALICE:2016kpq}. Besides, $SC^{v}(m, n)$ is independent on the symmetry plane correlations by design~\cite{Bilandzic:2013kga}.

Fig.~\ref{fig:sc_ALICE} (left) shows the centrality dependent symmetric cummulants $SC^{v}(4, 2)$ and $SC^{v}(3, 2)$ in 2.76 A TeV Pb--Pb collisions, measured from ALICE~\cite{ALICE:2016kpq} and calculated from the EKRT event-by-event hydrodynamics~\cite{Niemi:2015qia}. The positive values of $SC^{v}(4, 2)$ and negative values of $SC^{v}(3, 2)$ are consistent with the early observation from ATLAS~\cite{Aad:2015lwa}, which also illustrates that $v_{2}$ is anti-correlated with $v_{3}$, but correlated with $v_{4}$.  A comparison between the model calculations and the experimental data in Fig.~\ref{fig:sc_ALICE} also shows that,  although  hydrodynamics could successfully reproduce the integrated flow harmonics $v_{n}$, it can only qualitatively, but not quantitatively describe the correlations between these harmonics.

In Ref.~\cite{Zhu:2016puf}, the symmetric cumulants  $SC^{v}(m, n)$ and other related observables have been systematically calculated by the event-by-event viscous hydrodynamics VISH2+1 with a focus on investigating the influences from different initial conditions and QGP shear viscosity. Like the case of the early EKRT hydrodynamic simulations, all of these VISH2+1 simulations with MC-Glauber, MC-KLN and AMPT initial conditions could capture the sign and centrality dependence of $SC^{v}(4, 2)$ and $SC^{v}(3, 2)$, but not be able to archive a simultaneous quantitative descriptions of these two symmetric cumulants for all centrality intervals. Comparing with the individual flow harmonic $v_2$ and $v_3$, the symmetric cumulants $SC^{v}(4,2)$ and $SC^{v}(3,2)$ are more sensitive to the details of the theoretical calculations. Ref.~\cite{Zhu:2016puf} also predicted other
symmetric cumulants  $SC^{v}(5, 2)$, $SC^{v}(5, 3)$, and $SC^{v}(4, 3)$ and found that $v_2$ and $v_5$, $v_3$ and $v_5$ are correlated, $v_3$ and $v_4$ are anti-correlated for various centralities.

In order to get rid of the influences from individual flow harmonics, it was suggested to normalize $SC^{v}(m,n)$ by dividing the products $\left<v_m^2\right>\left<v_n^2\right>$~\cite{ALICE:2016kpq}. Fig.~\ref{fig:sc_ALICE} (right) and Fig.~\ref{fig:sc_hydro4} (a,b,c,g,h) plots the normalized symmetric cumulants $NSC^{v}(n,m)$ ($NSC^{v}(n,m)$ = $SC^{v}(n,m)$/$\left<v_n^2\right>\left<v_m^2\right>$) in 2.76 A TeV Pb--Pb collisions. $NSC^{v}(4,2)$ exhibits a clear sensitivity to the initial conditions and the $\eta/s(T)$ parameterizations, which  could provide additional constrains for the initial geometry and
the transport coefficients of the hot QCD matter. In contrast $NSC^{v}(3,2)$ is insensitive to the detailed setting of $\eta/s$ and the used initial conditions. Fig.~\ref{fig:sc_hydro4} also shows that the values of $NSC^{v}(3,2)$ is compatible to the ones of $NSC^{\varepsilon}(3,2)$ from the initial state due to the linear response of $v_2$ ($v_3$) to $\varepsilon_2$ ($\varepsilon_3$). Note that these different $NSC^{v}(3, 2)$ curves in Fig.~\ref{fig:sc_hydro4} (g) are almost overlap with each other, which also roughly fit the normalized ALICE data. In contrast, the predicted $NSC^{v}(4, 2)$, $NSC^{v}(5, 2)$, and $NSC^{v}(5, 3)$ are sensitive to both initial conditions and $\eta/s$. Due to the nonlinear hydrodynamic response,
$NSC^{v}(4, 3)$ does not necessarily follow the sign of $NSC^{\varepsilon}(4, 3)$ for some certain initial conditions.

In a recent work~\cite{Giacalone:2016afq}, the $NSC^{v}(m,n)$ are expressed in terms of symmetry plane correlations and moments of $v_{2}$ and $v_{3}$. Considering the relative flow fluctuations of $v_{3}$ is stronger than $v_{2}$, one expects smaller values for $NSC^{v}(5,2)$ compared to $NSC^{v}(5,3)$, as shown in Fig.~\ref{fig:sc_hydro4}.
On the other hand, it was predicted that $NSC^{v}(m,n)$ involving $v_{4}$ and $v_{5}$ increases with $\eta/s$ in the same way as the symmetry plane correlations~\cite{Teaney:2012ke, Teaney:2012gu}, which qualitatively agrees with the results in   Fig.~\ref{fig:sc_hydro4} from most central collisions to semi-peripheral collisions. \\

As discussed above, the low flow harmonics, $v_{2}$ or $v_{3}$, is mainly determined by a linear response to the initial eccentricity $\varepsilon_{2}$ or $\varepsilon_{3}$, while higher flow harmonics($v_{n}$ with $n>$ 3) not only contains the contributions  from the linear response of the corresponding $\varepsilon_{n}$, but also has additional contributions from lower order initial anisotropy coefficients. These additional contributions are usually called non-linear response of higher flow harmonics~\cite{Bhalerao:2014xra,Yan:2015jma}. In Ref.~\cite{Giacalone:2016afq}, it was proposed that a direct connection between symmetry plane correlations and the flow harmonic correlations $ NSC^{v}(m,n)$ could be built from the nonlinear hydrodynamic response of higher flow harmonics. Besides, the past hydrodynamic calculations have shown that the contributions of nonlinear response can explain the symmetry plane correlations and its centrality dependence~\cite{Yan:2015jma,Qian:2016pau}. Recently, the proposed nonlinear hydrodynamic coefficient~\cite{Yan:2015jma} has been systematically studied and measured~\cite{Qian:2016pau,ALICE:NLR,CMS:NLR}, which could be used to further constrain the initial conditions and $\eta/s$, and to provide a better understand of the correlations between different flow harmonics.


\section{Correlations and Collective flow in small systems}

\subsection{p--Pb collisions at $\sqrt{s_{\rm NN}}=$ 5.02 TeV }

\begin{figure*}[t]
\centering
\includegraphics[width=0.8\textwidth]{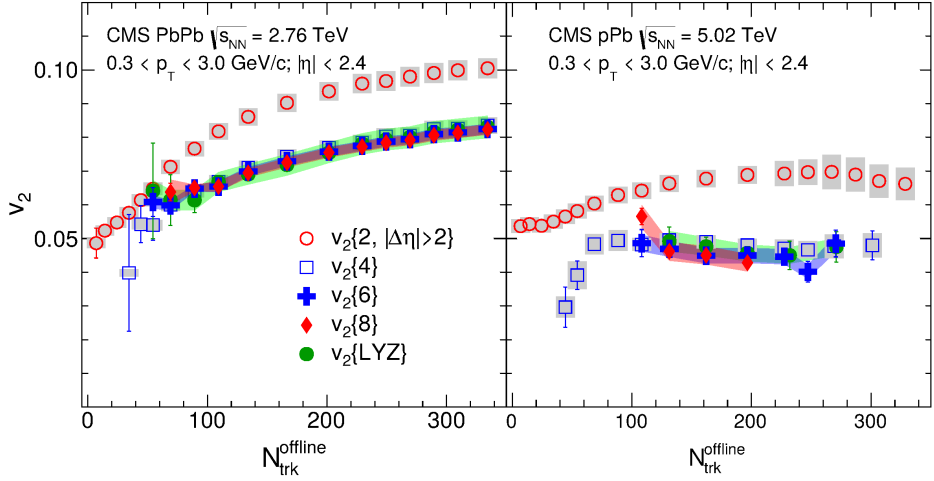}
\caption{(Color online) Multiplicity dependence of $v_2$, obtained from Fourier decomposition of 2-particle azimuthal correlations, from multi-particle cumulants, and via LYZ method, in Pb--Pb collisions at $\sqrt{s_{\rm NN}}=$ 2.76 TeV (left) and p--Pb collisions at $\sqrt{s_{\rm NN}}=$ 5.02 TeV (right)~\cite{Khachatryan:2015waa}.}
\label{fig:CMSpPb}
\end{figure*}

High energy proton-lead (p-Pb) collisions at the LHC was originally aimed to study the cold nuclear matter effects and provide the corresponding reference data for Pb--Pb collisions at the LHC.  However, lots of unexpected collective phenomena have been observed in experiments. For example, the measured two particle correlations showed a symmetric double ridge structure on both near-and away-side in high multiplicity p--Pb collisions at $\sqrt{s_{\rm NN}}=$ 5.02 TeV~\cite{CMS:2012qk,Abelev:2012ola,Aad:2013fja,Khachatryan:2015waa}.
Besides, negative 4- and 8-particle cumulants and positive 6-particle cumulants have been observed in the high multiplicity events~\cite{Aad:2013fja,Abelev:2014mda,Khachatryan:2015waa}. In particular, all the multi-particle cumulants (including 4-, 6- and 8-particles cumulants) are compatible to the ones obtained from all-particle correlations with Lee-Yang Zero's method, which corresponds to $v_{2}\{4\} \approx v_{2}\{6\} \approx v_{2}\{8\} \approx v_{2}\{{\rm LYZ}\}$~\cite{Khachatryan:2015waa}), as shown in Fig.~\ref{fig:CMSpPb} (This observation has also been confirmed by the later ATLAS~\cite{Aad:2013fja} and ALICE Collaborations~\cite{Abelev:2014mda} measurements).
Meanwhile, the obtained $v_{2}$ from two or four-particle cumulants are comparable to the ones from Pb--Pb collisions at 2.76 TeV~\cite{Aad:2013fja,Chatrchyan:2013nka,ABELEV:2013wsa,Khachatryan:2015waa}. Recently,
the ALICE collaboration has extended the investigated of anisotropic collectivity via azimuthal correlations of identified hadrons~\cite{ABELEV:2013wsa,Khachatryan:2014jra}. A typical mass-ordering feature among the $v_{2}$ of pions, kaons and protons is observed in high multiplicity p-Pb collisions~\cite{ABELEV:2013wsa}. Similarly, the CMS Collaboration found a $v_2$ mass-ordering between ${\rm K_{S}^{0}}$  and $\Lambda(\overline{\Lambda})$~\cite{Khachatryan:2014jra}.

There are many theoretical efforts attempt to provide explanation for the flow-like behavior of the p--Pb collisions. In general they can be divided into two big categories that doesn't involve the final-state evolution of the medium but only account for initial-state effects~\cite{Dusling:2012iga,Dusling:2012cg,Dusling:2012wy,
Dusling:2013qoz,Dusling:2014oha,Kovner:2012jm,Dumitru:2014dra,Dumitru:2014vka,Noronha:2014vva}, and that include the final-state interactions, such as the hydrodynamics or kinetic model description~\cite{Bozek:2011if,Bozek:2012gr,Bozek:2011if,Bozek:2013ska,Bzdak:2013zma,Qin:2013bha,
Werner:2013ipa,Schenke:2014zha,Bzdak:2014dia,Ma:2014pva,Bozek:2015swa,Koop:2015wea,Li:2016ubw,Zhou:2015iba}.
In this section, we will focus on reviewing the hydrodynamic calculations as well as the kinetic model investigations on the flow-like signals in the small p--Pb systems.

\begin{figure*}[t]
\includegraphics[angle=0,width=0.49 \textwidth, height=6.2cm]{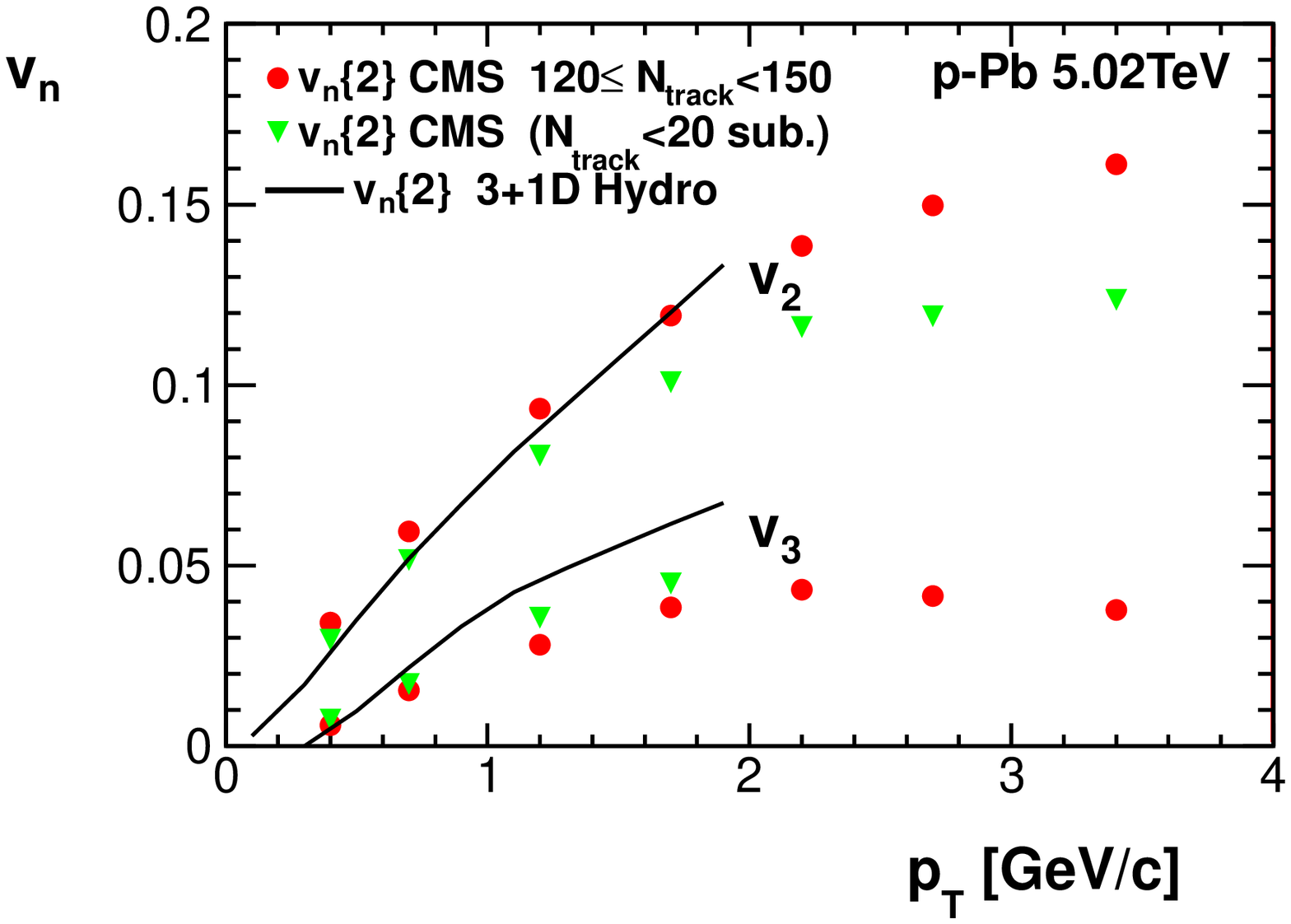}
\includegraphics[angle=0,width=0.49 \textwidth, height=6.2cm]{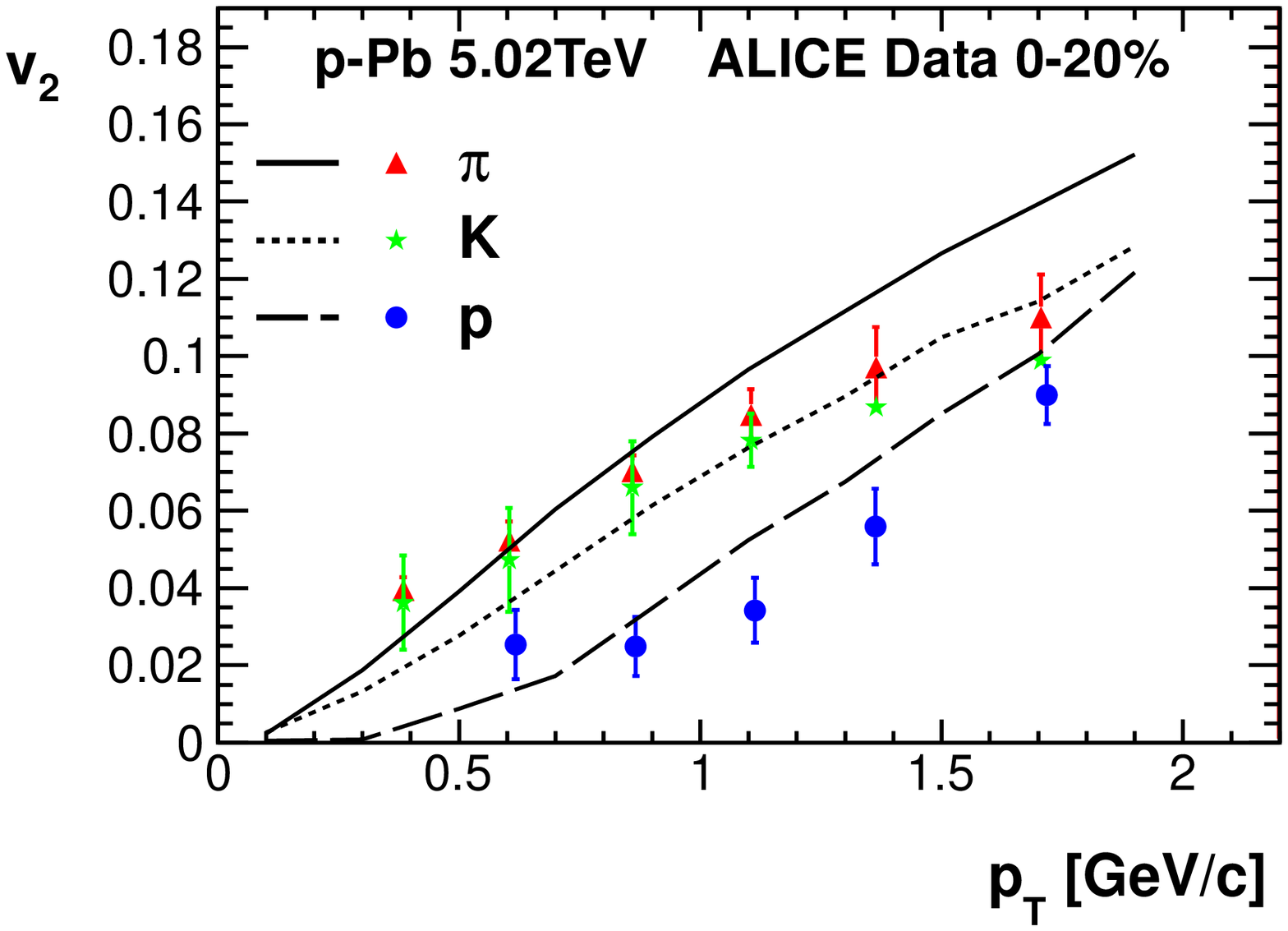}
\caption{(Color online) The hydrodynamic calculations of the elliptic and triangular flow coefficient of all charged particles (left panel) and elliptic flow of identified hadrons (right panel) in p-Pb collisions at $\sqrt{s_{\rm NN}}=$ 5.02 TeV~\cite{Bozek:2013ska}, together with a comparison with the CMS~\cite{Chatrchyan:2013nka} and ALICE data
data~\cite{ABELEV:2013wsa}.
\label{fig:v23}}
\end{figure*}

\vspace{0.2cm}
\underline{\emph{Results from hydrodynamic simulations}}:
\vspace{0.10cm}

Hydrodynamics is a useful tool to simulate the collective expansion of the created systems and quantitatively study and predict the final flow observable. Recently, the holographic duality calculations have shown that the size of the produced droplet is $\sim 1/T_{eff}$~\cite{Chesler:2015bba,Chesler:2016ceu}, which indicate that hydrodynamics is possibly applicable for the small systems created in the high energy p--Pb and p--p collisions. Using 3+1-d hydrodynamic or hybrid model simulations, different groups has systematically studied the the multiplicities, mean $p_T$, final state correlations and related flow data
in p--Pb collisions at $\sqrt{s_{\rm NN}}=$ 5.02 TeV~\cite{Bozek:2012gr,Bozek:2011if,Bozek:2013ska,Bzdak:2013zma,Qin:2013bha,Werner:2013ipa,Schenke:2014zha}.
In general, these hydrodynamic calculations could semi-quantitatively described these different soft hadron data, which support the observation of collective flow in experiments of high energy p--Pb collisions.

Fig.13 (left) presents the hydrodynamic calculations for flow coefficients $v_2$ and $v_3$ of all charged hadrons in high multiplicity p--Pb collisions, which give a roughly fit of the data from the CMS collaborations~\cite{Bozek:2013ska}.  It was also found such fluid evolution also develop the radial flow, which
leads to a flatter transverse momentum spectra for various hadron species. As shown in Ref~\cite{Bozek:2013ska}, the average transverse momentum of the identified hadrons in p--Pb collisions can be consistently fitted by the hydrodynamic simulations. In contrast, the HIJING model without any collective expansion fails to describe the data. In the hydrodynamic language, the interaction between radial and elliptic flow re-distribute the total momentum anisotropy to various hadron species, leading to a mass ordering of the flow harmonics. Fig.13 (right) shows that the hydrodynamic simulations roughly reproduce the $v_2$ mass-ordering of  pions. kaons and protons. Note that, other hydrodynamic calculations with different initial conditions and transport coefficients also obtained similar results. For details, please refer to~\cite{Bzdak:2013zma,Qin:2013bha,Werner:2013ipa,Schenke:2014zha}.

Ref.~\cite{Bozek:2013uha} has shown that, in order to reproduce the multiplicity distribution of p--Pb collisions using the
hydrodynamic calculations with Glauber initial conditions, the implementation of  additional negative binomial fluctuations
are necessary. Correspondingly, initial eccentricities are also modified, which leads to a simultaneous fit of the $v_2\{2\}$ and  $v_2\{4\}$ data.  In contrast, the early IP-glasma initial condition generates the initial energy distributions with an imprinted spherical shape of protons, which yields a very small $v_2$ for the p--Pb collision systems~\cite{Schenke:2014zha}. This motivates the recent investigations of the proton structure within the saturation framework, which indicates that the shape of the protons also fluctuate event-by-event~\cite{Mantysaari:2016ykx,Mantysaari:2016jaz}.

Note that the flow-like signals have also been observed in d--Au and $^3\mathrm{He}$--Au collisions at RHIC. Compared to the p--A collisions at the LHC, the d--Au and $^3\mathrm{He}$--Au collisions provide controlled initial geometry deformations, which are less sensitive to the details of initial state models and are helpful to check the hydrodynamic caculations. Recently, the STAR and PHENIX collaboration has measured the elliptic flow $v_2$ in d--Au collisions at $\sqrt{s_{\rm NN}}=$ 200 GeV and the elliptic and triangular flow $v_2$ and $v_3$  in $^3\mathrm{He}$--Au collisions at $\sqrt{s_{\rm NN}}=$ 200 GeV ~\cite{Adare:2013piz,Adare:2014keg,Adamczyk:2015xjc,Adare:2015ctn}. The hydrodynamic calculations from different groups, using various initial conditions and the QGP shear viscosity, roughly described these extracted flow data. It was also found that $v_2$ and $v_3$ follows $\varepsilon_2$ and $\varepsilon_3$ from the initial state, which give a support for the collective expansion in these small systems created at RHIC~\cite{Bzdak:2013zma,Qin:2013bha,Koop:2015trj,Bozek:2015qpa,Romatschke:2015gxa}
.

Compared with the case in Pb--Pb collisions, the initial sizes of the created systems in p--Pb collisions are much smaller. The subsequent collective expansion is expected to enlarge the size of the fireball, where the corresponding radii at the freeze-out can be measured by the Hanbury-Brown Twiss (HBT) correlations. In Ref~\cite{Adam:2015pya}, the ALICE collaboration has measured the three-dimensional pion femtoscopic radii in p--Pb collisions at $\sqrt{s_{\rm NN}}=$ 5.02 TeV, which showed that the size of the p--Pb systems is in between the ones obtained from p--p collisions and peripheral Pb--Pb collisions.  In general, the hydrodynamic calculations could roughly describe the HBT measurements, while the quantitative values from different model calculations are sensitive to the initial conditions and the imprinted initial sizes of the created fireball~\cite{Romatschke:2015gxa,Bozek:2014fxa,Shapoval:2013jca}.

In~\cite{Niemi:2014wta}, the validity of hydrodynamics for large Pb--Pb and small p--Pb systems at the LHC has been evaluated through tracing the space time evolution of the Knudsen number. It was found for Pb--Pb collisions, hydrodynamic simulations with $\eta/s \sim 1/4\pi$ are always within the validity regime with the Knudsen numbers well below one.  However, the related simulations for smaller p--A systems shows that the hydrodynamic descriptions has broken down at the $T_{dec} = 100 \ \mathrm{MeV}$ freeze-out boundary even using a minimum QGP
shear viscosity as a input. Although such investigations will not preclude the collective flow and final state
interactions, it is worthwhile to explore the physics of the small p--Pb systems within other framework beyond hydrodynamics.

\begin{figure*}[t]
\centering
\includegraphics[width=0.6\textwidth,height=5.6cm]{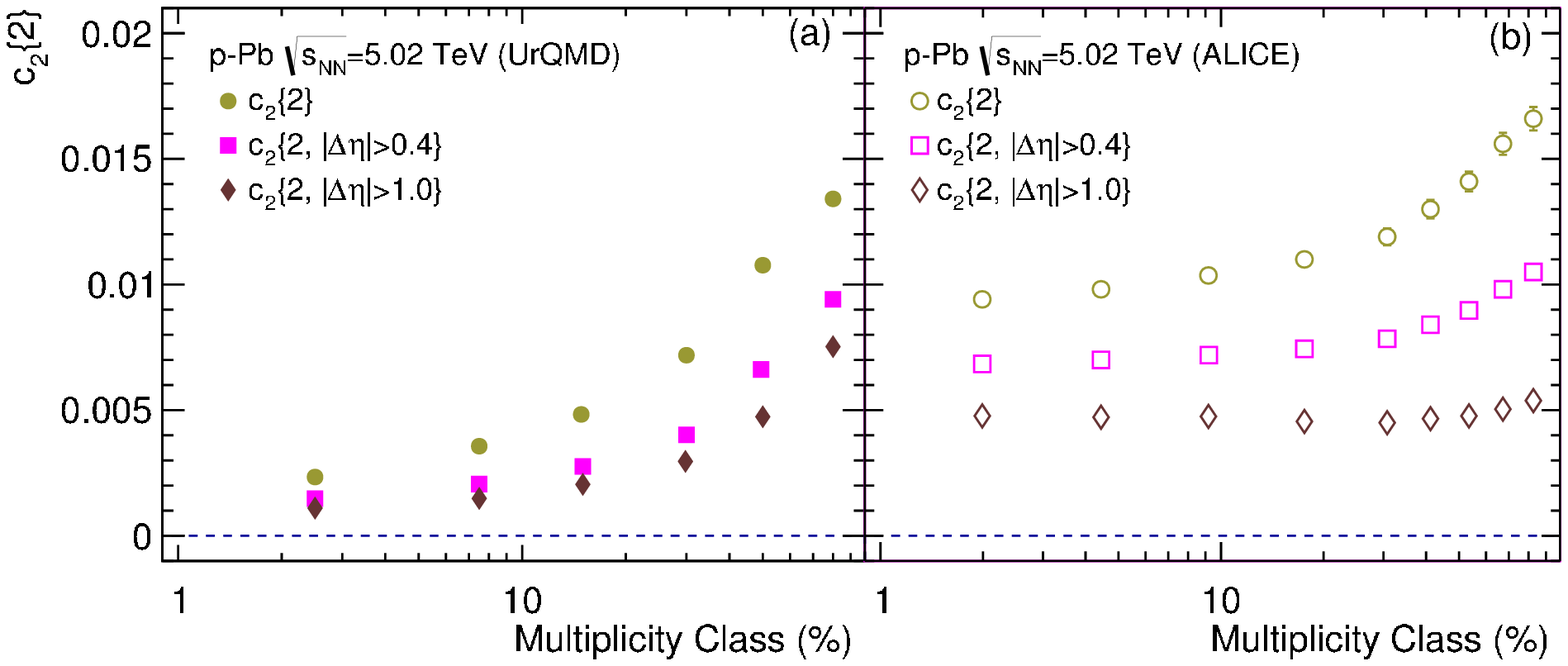}
\includegraphics[width=0.35\textwidth,height=5.6cm]{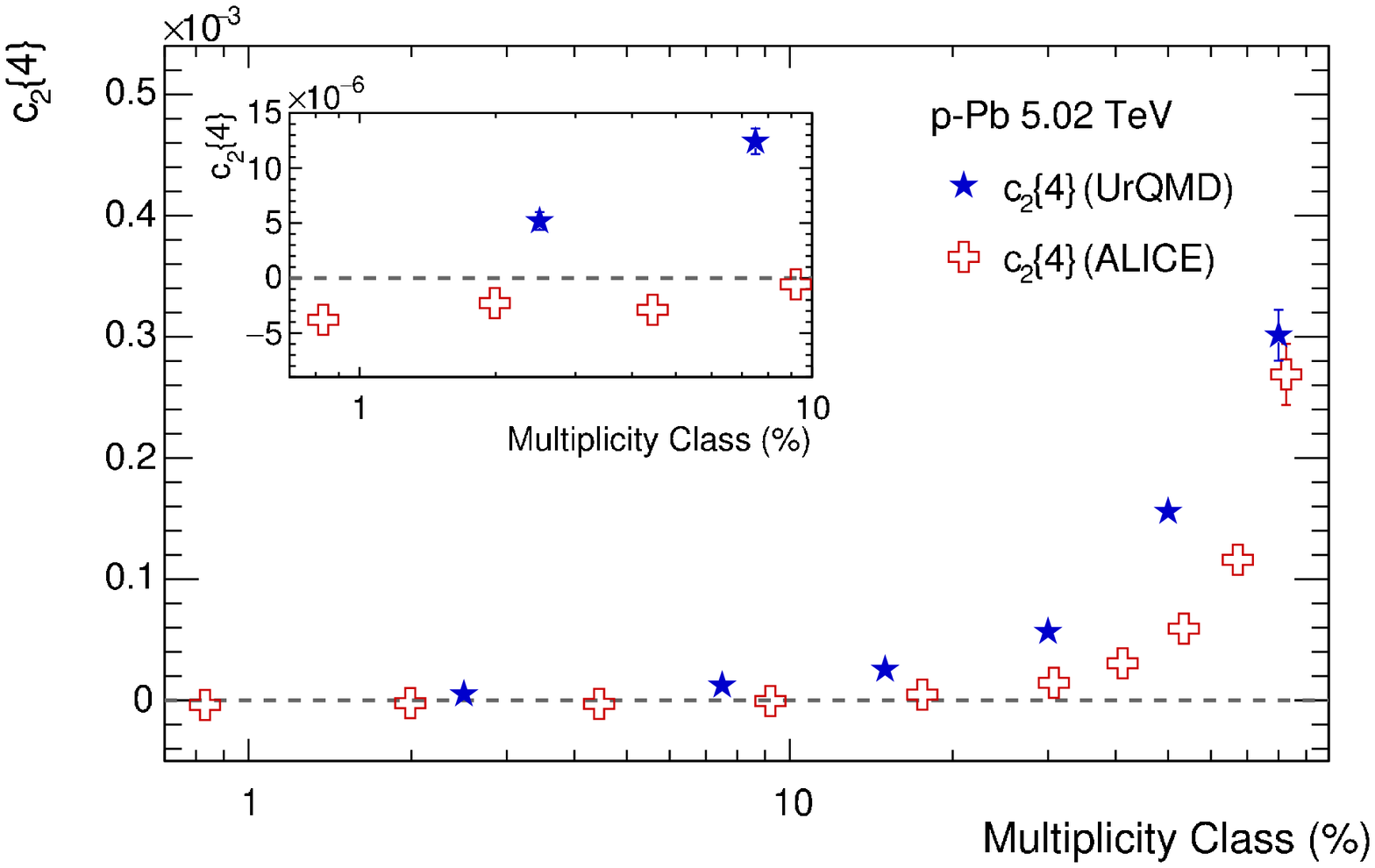}
\caption{(Color online) Centrality dependence of $c_{2}\{2\}$ (left) and $c_{2}\{4\}$ (right), calculated from UrQMD~\cite{Zhou:2015iba} and measured by ALICE~\cite{Abelev:2014mda}.}
\label{figure:c224}
\end{figure*}

\vspace{0.2cm}
\underline{\emph{Results from other approaches}}:
\vspace{0.10cm}

Without the final state interactions, the long range rapidity correlations in high energy
p--p and p--Pb collisions have been calculated with the framework of Color Glass Condensate (CGC),
which shows a good agreement with the di-hadron data from the CMS, ATLAS and ALICE~\cite{Dusling:2012iga,Dusling:2012cg,Dusling:2012wy,Dusling:2013qoz}.  However the odd harmonics data
disfavor this early CGC calculations without the rescattering contributions~\cite{Dusling:2014oha}. Without a proper hadronization procedure, such calculations can
also not predict the flow data of the identified hadrons.
Recently, it was proposed that a presence of the colored domains inside the proton and the nucleus breaks
rotational invariance, which helps to generate elliptic and triangular flow during the scattrings between
a dilute projectile of valence quarks and the
nucleus~\cite{Kovner:2012jm,Dumitru:2014dra,Dumitru:2014vka,Noronha:2014vva}.
An alternative approach is the classical Yang-Mills simulations, which treat
both proton and nucleus as dense QCD objects with high gluon
occupancy and are more appropriate to describe the early
time evolution of the created p--Pb systems in the high multiplicity events.
Within such framework,  Schenke and his collaborators have calculated the
single and double inclusive gluon distributions and extracted the associated $p_T$ dependent
elliptic and triangular flow of gluons in high energy p--A collisions~\cite{Schenke:2015aqa}. They found that the final state
effects in the classical Yang-Mills evolution build up a non-zero triangular flow, but only slightly
modify the large elliptic flow of gluons created from the initial state~\cite{Schenke:2015aqa}. Although this investigation
only focus on the flow anisotropy of gluons,
the obtained large value of $v_2$ and $v_3$ indicate such pre-equilibrium dynamics
should be combined with the model calculations of the final state interactions, such as hydrodynamics
or the Boltzmann simulations.

The flow signals in the p--Pb collisions have also been investigated within the framework of multiphase transport model (AMPT)~\cite{Bzdak:2014dia,Ma:2014pva,Bozek:2015swa,Koop:2015wea,Li:2016ubw}. With a tuned cross-sections within the allowed range $\sigma\sim 1.5-3 \ \mathrm{mb}$,  AMPT nicely fit the two particle correlations and the extracted $v_2$ and $v_3$ coefficients in high energy p--Pb collisions~\cite{Bzdak:2014dia,Ma:2014pva}. Ref.~\cite{Bzdak:2014dia,Li:2016ubw} has shown that AMPT generates a mass-ordering of $v_2$ and $v_3$  for various hadron species with the coalescence process tuning on. It was also surprisingly observed that the collective behavior in AMPT is built up by a small amount of interactions, where each parton undergoes two collisions on average. The escape mechanism prosed in ~\cite{Li:2016ubw, Li:2016flp} seems to be responsible for the anisotropy buildup in AMPT, but is dramatically different from the traditional flow development picture of hydrodynamics due to the strong interactions.

\begin{figure*}[t]
\centering
\includegraphics[width=0.4\textwidth]{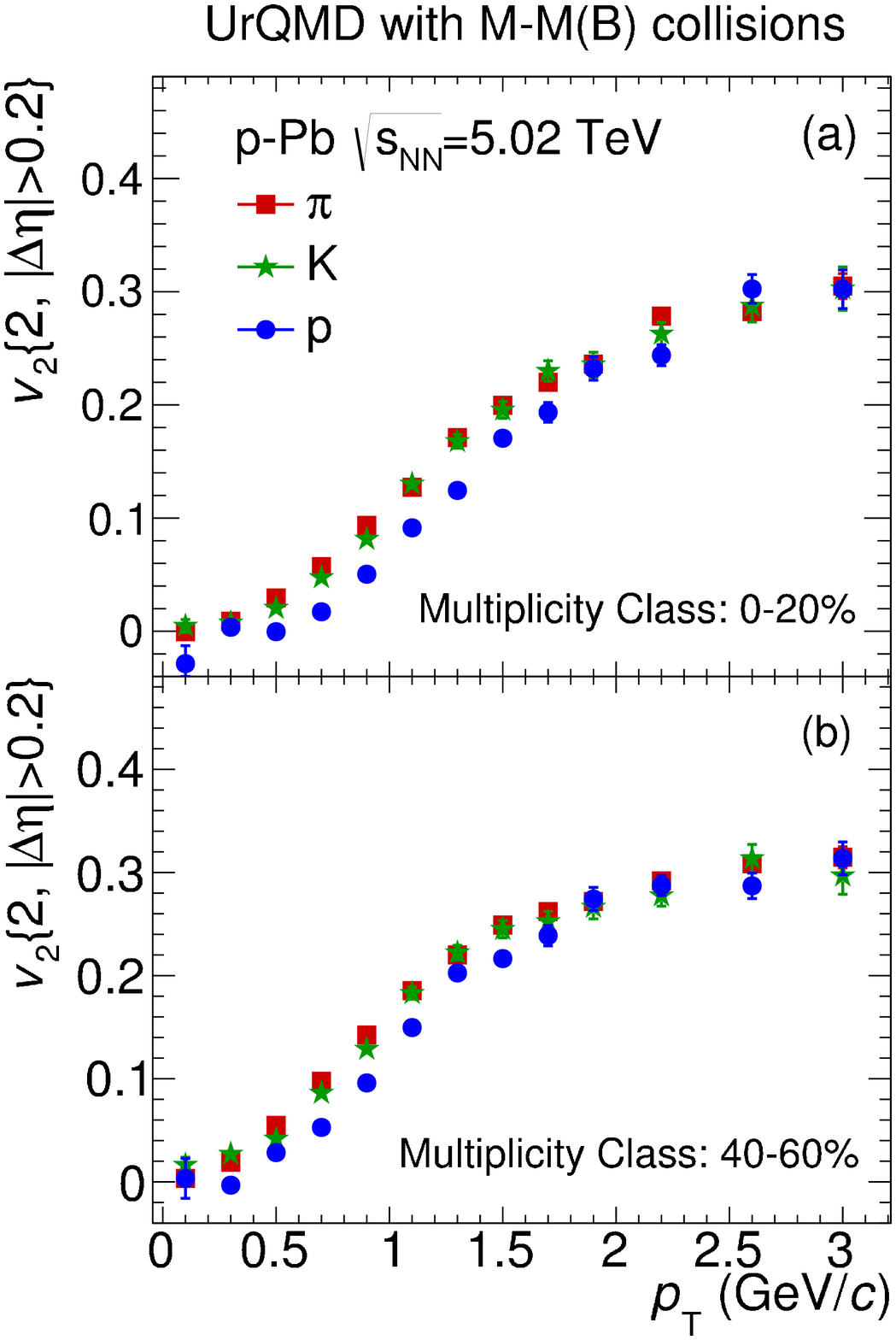}
\includegraphics[width=0.4\textwidth]{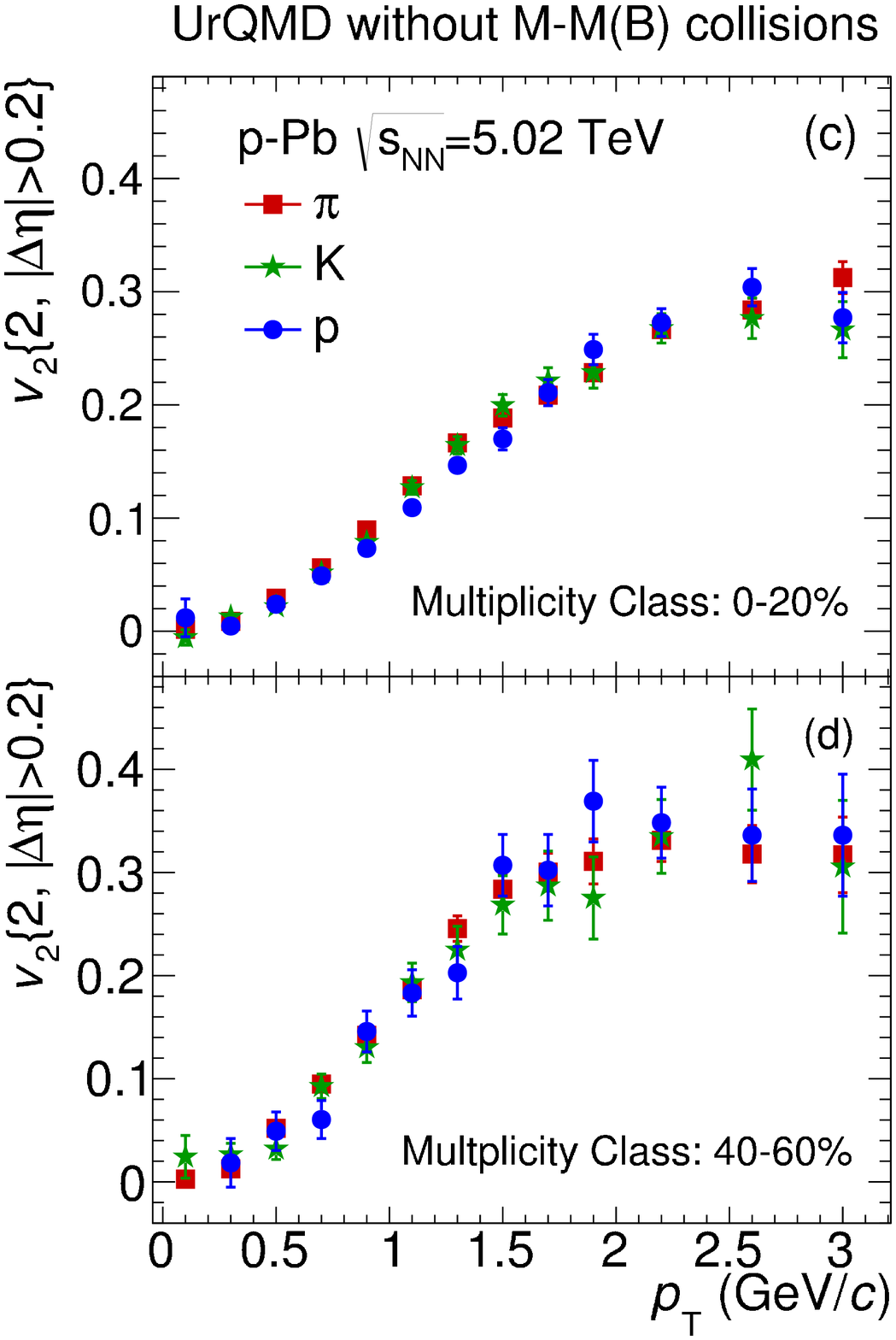}
\caption{(Color online) $v_{2}(\it{p}_{\rm T})$ of pions, kaons and protons in p--Pb collisions at $\sqrt{s_{_{\rm NN}}} =$ 5.02 TeV, calculated from {\tt UrQMD} with and without M-M and M-B collisions~\cite{Zhou:2015iba}.}
\label{fig:pidv2Gap02}
\end{figure*}

With an assumption that the high energy p--Pb collisions do not reach the threshold to create the QGP, but only produce pure hadronic systems, Ref~\cite{Zhou:2015iba} systematically investigated the 2 and 4 particle correlations of all charged and identified hadrons, using the hadron cascade model Ultra-relativistic Quantum Molecular Dynamics (UrQMD )~\cite{Bass:1998ca,Bleicher:1999xi,Petersen:2008kb}. Fig.~14 shows the two and four -particle cumulants $c_{2}\{2\}$  and $c_{2}\{4\}$ of all charged hadrons, calculated from UrQMD and measured from ALICE. In general, $c_{2}\{2\}$ decreases with the increase of the pseudorapidity gap, which is agree with the expectation of suppressing the non-flow effects with a large pseudorapidity gap. However, UrQMD still presents a strong centrality dependence of $c_{2}\{2\}$ for $|\Delta \eta|>1.0$, which indicates that the remaining non-flow effects are still strong there. In Fig.~\ref{figure:c224} (right), the $c_{2}\{4\}$ from ALICE exhibits a transition from positive to negative values, which indicate the creation of flow-dominated systems for the high multiplicity events. In contrast, $c_{2}\{4\}$ from UrQMD simulations keeps positive for all multiplicity classes, which illustrates that the p--Pb systems created by UrQMD are non-flow dominated.

However, the generally believed collective expansion feature, the mass-ordering of $v_{2}(p_{\rm T})$, are reproduced in
the UrQMD simulations. Fig.~\ref{fig:pidv2Gap02} shows that these high multiplicity events from UrQMD
present a clear $v_2$ mass-ordering among pions, kaons and protons, which are qualitatively agrees with the corresponding ALICE measurement~\cite{ABELEV:2013wsa}. In  UrQMD, the meson-baryon (M-B) cross sections from AQM are about 50\% larger than the meson-meson (M-M) ones, which leads to the $v_{2}$ splitting between mesons and baryons in the UrQMD simulations. Fig.~\ref{fig:pidv2Gap02} also shows, after switching off the M-B and M-M interaction channels, the characteristic feature of $v_{2}$ mass-ordering disappears. Therefore, even without enough flow generation, the hadronic interactions still lead to a $v_{2}$ mass-ordering feature for a hadronic p--Pb system.

In Ref~\cite{Romatschke:2015dha}, the created p--Pb systems are described by non-interacting free-streaming particles, following with a harmonization procedure and a hadronic cascade evolution. Such non-hydrodynamic simulations showed,
although the elliptic flow are under-predicted, the triangular and quadrupolar flow are raised by the free-streaming
evolution, which are comparable to the ones obtained from the hydrodynamic simulations. Meanwhile, the $v_n$ mass-orderings
among pions, kaons and protons have also been observed in such non-hydrodynamic p--Pb systems due to
the hadronic interactions during the late evolution.

%

\subsection{p--p collisions at $\sqrt{s_{\rm NN}}=$ 7 TeV and 13 TeV}
Like the case for high energy p--Pb collisions, the long-range two-particle azimuthal correlations with a large pseudo-rapidity separation have also been observed in high-multiplicity p--p collisions at the LHC, which provides new insights for the novel dynamics of the small QCD systems~\cite{Khachatryan:2010gv,Li:2012hc,Khachatryan:2015lva,Aad:2015gqa,Khachatryan:2016txc}.
For p--Pb collisions at $\sqrt{s_{\rm NN}}=$ 5.02 TeV, the extensive measurements of the 2 particle and multi-particle correlations, extracted flow harmonics for all charged and identified hadrons, as well as the supportive hydrodynamic calculations strongly indicates that collective expansion has been developed in the small p--Pb systems. However, for high-energy p--p collisions at the LHC, the nature of the observed long-range correlation is still an open question (For different theoretical interpretations, please refer to~\cite{Dusling:2013qoz,Dusling:2012iga,Dumitru:2010iy,Levin:2011fb,Tribedy:2011aa,
Bozek:2010pb,Bzdak:2013zma,Werner:2010ss,Schenke:2014zha,Schenke:2016lrs,Dusling:2015gta}).

Recently, the ATLAS Collaboration has measured the Fourier coefficients $v_{n}$ in p--p collisions at $\sqrt{s_{\rm NN}}=$ 13 TeV, using the two-particle correlations as a function of the relative azimuthal-angle and pseudo-rapidity~\cite{Aad:2015gqa}. It was found that the extracted $v_{2}$ is approximately a constant as a function of multiplicity and its $p_{\rm T}$ dependence is very similar to the one measured in p--Pb and Pb--Pb collisions~\cite{Aad:2015gqa}.
The CMS collaboration further measured the $v_n$ coefficients for all charged hadrons, as well as for $K_S^0$ and $\Lambda/ \overline{\Lambda}$ in p--p collisions at $\sqrt{s_{\rm NN}}=$ 5, 7 and 13 TeV, which
observed a clear $v_2$ mass-ordering among all charged hadrons, $K_S^0$ and $\Lambda/ \overline{\Lambda}$~\cite{Khachatryan:2016txc}.
Furthermore, the CMS collaboration has measured the multi-particle cumulants, the key observable to probe the anisotropic collectivity. A negative sign of $c_{2}\{4\}$ and a positive sign of $c_{2}\{6\}$ appeared in the high multiplicity p--p collisions at $\sqrt{s_{\rm NN}}=$  13 TeV ~\cite{Khachatryan:2016txc}, which seems to indicate the development of anisotropic collectivity in high energy p--p collisions. However, the ATLAS Collaboration reported in Hard Probe 2016 conference that the multiplicity fluctuations could significantly bias the measurements of multi-particle cumulants~\cite{ATLAS:HP2016}, which indicates that non-flow might mimic the flow signal by pushing the $c_2\{4\}$ to negative values. In order to avoid the bias from multiplicity fluctuations, the so-called ``Method1'', which using the same multiplicity selection for the calculations of cumulants and $N_{\rm trk}$, is applied. The obtained $c_{2}\{4\}$, which is less affected by multiplicity fluctuations, does not show  negative sign for the multiplicity regions where negative values of $c_{2}\{4\}$ was reported by CMS.

For small systems, it is also very important to address and evaluate the non-flow effects. Generally, the multi-particle cumulants, e.g. $c_2\{4\}$, are able to suppress the non-flow of two-particle correlations in traditional Au+Au or Pb+Pb collisions. However, the non-flow contributions to the multi-particle correlations are still remained and might play an non-negligible role in the small  p--p collision systems. Recently, the ALICE and ATLAS Collaborations have proposed new 4-particle cumulant methods with $|\Delta\eta|$ gap separation, using 2- or 3-subevents~\cite{ALICE:QM2017,ATLAS:QM2017}. By selecting particles from different regions separated by a $|\Delta\eta|$ gap, it is possible to further suppress the non-flow contributions in the multi-particle cumulants. This has been verified in the PYTHIA simulations~\cite{Jia:2017hbm}.  The preliminary measurements in p--p collisions at 13 TeV, reported in QM2017~\cite{ALICE:QM2017,ATLAS:QM2017}, have shown that the non-flow effects are suppressed with these new 4-particle cumulant methods. A negative sign of the 4-particle cumulant was observed by ATLAS collaboration after implementing the 3-subevent method, while ALICE has not confirm the negative sign of $c_2\{4\}$ with a $|\Delta\eta|$ gap separation due to the limited statistics and relatively smaller acceptance.

Besides the multi-particle cumulants for single flow harmonics, the CMS Collaboration also measured the symmetric cumulants SC(m,n) and normalized symmetric cumulants NSC(m,n) in p--p, p--Pb and Pb--Pb collisions~\cite{CMS:QM2017}. It was found that the normalized NSC(3,2) are similar in p--Pb and Pb--Pb collisions, indicating that these two systems present similar  initial state fluctuation patterns for the correlations between $\varepsilon_2$ and $\varepsilon_3$. While, the normalized NSC(4,2) shows certain orderings for the p--p, p--Pb and Pb--Pb collision systems, which may associates with the different non-linear response and non-flow effects between the large and small systems.

In short, these recent measurements in p--p collisions $\sqrt{s_{\rm NN}}=$  13 TeV are aimed to evaluate whether or not collective flow has been created in high multiplicity p--p collisions. Future investigations, from both experimental and theoretical sides, are very crucial to further address this question and for a deep understanding of the underline physics
in the small collision systems.

\section{Summary}
In this paper, we briefly reviewed the collective flow and hydrodynamics in large and small systems at the LHC.
One of the important messages we would like to convey to readers was that hydrodynamics and hybrid models are
important and useful tools to study various flow observables in high energy nucleus-nucleus and nucleus-nucleons collisions.
With a properly chosen initial condition and well tuned QGP transport coefficients, hydrodynamics and hybrid models can quantitatively describe the flow harmonics coefficients $v_n$ of all charged hadrons and make very nice predictions for the flow data of identified hadrons. The massive-data fitting  of the flow harmonics and other related soft hadron data, using the sophisticated hybrid model simulations, have extracted the functions of the temperature-dependent QGP shear and bulk viscosities at the LHC, which demonstrated that the created QGP is an almost perfect fluid with very small shear viscosity close to the KSS bound.

For some flow observables in the high energy Pb--Pb collisions, e.g. the event plane correlations, the correlations between different flow harmonics, etc., hydrodynamic and hybrid models can qualitatively, but not quantitatively, describe the data.
However, such qualitatively descriptions can still be considered as a success of the hydrodynamics, considering that the initial state fluctuations contain different intrinsic patens from the ones extracted from final state correlations.  The succeeding hydrodynamic evolution drastically change some of these initial state correlations, even the signs, making a quantitatively description of the data. On the other hand, these flow data are more sensitive to the details of theoretical model calculations. A further study of these flow observables could reveals more information on the initial state fluctuations, non-linear hydrodynamic response and etc., which could also help us to further constraint the initial state models and to  precisely extract the QGP transport coefficients in the future.

As a hot research topic, the flow-like signals in high energy p--Pb and p--p collisions at the LHC have been widely investigated in both experiment and theory. For the high multiplicity p--Pb collisions, the observation of the changing sign of the 4 particle cumulants, the $v_2$ mass orderings, and the supportive calculations from hydrodynamics, etc., strongly indicated the development of collective expansion in the small p--Pb systems. For the high energy p--p collisions, some similar results, but with smaller magnitudes, have been observed for many flow-like observables.  Although these measurements may also associated with the collective expansion, more detailed investigations are still needed to further understand of the physics in the small p--p systems.\\

\noindent\textbf{\emph{Acknowledgments}}

This work is supported by the NSFC and the MOST under grant Nos.11435001, 11675004 and 2015CB856900 and by the Danish Council for Independent Research, Natural Sciences, and the Danish National Research Foundation (Danmarks Grundforskningsfond).

\bibliography{bibliography}

\end{document}